\documentclass[12pt]{article}
\usepackage{aaspp4}
\usepackage{flushrt}
\journalid{337}{15 January 1989}
\articleid{11}{14}
\usepackage{epsfig}

\def\etal{{et al.}}
\begin{document}
\setcounter{figure}{0}
\title{Proper Motions of Dwarf Spheroidal Galaxies from \textit{Hubble
Space Telescope} Imaging.  I: Method and a Preliminary Measurement for
Fornax.\footnote{Based on observations with NASA/ESA \textit{Hubble Space
Telescope}, obtained at the Space Telescope Science Institute, which is
operated by the Association of Universities for Research in Astronomy,
Inc., under NASA contract NAS 5-26555.}}

\author{Slawomir Piatek} \affil{Dept. of Physics, New Jersey Institute
of Technology,
Newark, NJ 07102 \\ E-mail address: piatek@physics.rutgers.edu}

\author{Carlton Pryor}
\affil{Dept. of Physics and Astronomy, Rutgers, the State University
of New Jersey, 136~Frelinghuysen Rd., Piscataway, NJ 08854--8019 \\
E-mail address: pryor@physics.rutgers.edu}

\author{Edward W.\ Olszewski}
\affil{Steward Observatory, University of Arizona,
    Tucson, AZ 85721 \\ E-mail address: eolszewski@as.arizona.edu}

\author{Hugh C.\ Harris}
\affil{US Naval Observatory, Flagstaff Station, P. O. Box 1149,
Flagstaff, AZ 86002-1149 \\ E-mail address: hch@nofs.navy.mil}

\author{Mario Mateo} \affil{Dept. of Astronomy, University of
Michigan, 830 Dennison Building, Ann Arbor, MI 48109-1090 \\
E-mail address: mateo@astro.lsa.umich.edu}

\author{Dante Minniti}
\affil{Department of Astronomy, Pontificia Universidad Cat\'{o}lica,
Av.~Vicu\~{n}a Mackenna 4860, Santiago 22, Chile \\
E-mail address: dante@astro.puc.cl}

\author{David G. Monet}
\affil{US Naval Observatory, Flagstaff Station, P. O. Box 1149, 
Flagstaff, AZ 86002-1149 \\
E-mail: dgm@nofs.navy.mil}

\author{Heather Morrison} \affil{Department of Astronomy and
Department of Physics, Case Western Reserve University, 10900 
Euclid Avenue, Cleveland, OH 44106-7215 \\
E-mail address: heather@vegemite.astr.cwru.edu}

\author{Christopher G.\ Tinney}
\affil{Anglo-Australian Observatory, PO Box 296, Epping, 1710,
Australia \\ E-mail address: cgt@aaoepp.aao.gov.au}

\begin{abstract}

	This article presents and discusses a method for measuring the
proper motions of the Galactic dwarf spheroidal galaxies using images
taken with the Hubble Space Telescope.  The method involves fitting an
effective point spread function to the image of a star or quasi-stellar
object to determine its centroid with an accuracy of about 0.005~pixel
(0.25~milliarcseconds) --- an accuracy sufficient to measure the proper
motion of a dwarf spheroidal galaxy using images separated by just a
few years.  The data consist of images, dithered to reduce the effects
of undersampling, taken at multiple epochs with the Space Telescope
Imaging Spectrograph or the Wide Field Planetary Camera.  The science
fields are in the directions of the Carina, Fornax, Sculptor, and Ursa
Minor dwarf spheroidal galaxies and each has at least one quasi-stellar
object whose identity has been established by other studies.  The rate
of change with time of the centroids of the stars of the dwarf
spheroidal with respect to the centroid of the quasi-stellar object is
the proper motion.

Four independent preliminary measurements of the proper motion of
Fornax for three fields agree within their uncertainties.  The weighted
average of these measurements is $\mu_{\alpha} = 49 \pm
13$~milliarcseconds~century$^{-1}$ and  $\mu_{\delta} = -59 \pm
13$~milliarcseconds~century$^{-1}$.  The Galactocentric velocity
derived from the proper motion implies that Fornax is near
perigalacticon, may not be bound to the Milky Way, and is not a member
of any of the proposed streams of galaxies and globular clusters in the
Galactic halo.  If Fornax is bound, the Milky Way must have a mass of
at least $(1.6 \pm 0.8) \times 10^{12}~\mathcal{M}_{\odot}$.

\end{abstract}


\section{Introduction}
\label{intro}

	There are nine known dwarf spheroidal (dSph, hereafter)
galaxies in relative proximity to the Milky Way, (see Mateo 1998 and
van den Bergh 2000 for reviews).  The nearest, Sagittarius, is
only about 16~kpc from the Galactic center and it is moving on a polar
orbit.  The most distant dSph, Leo~I, is 250~kpc from the Galactic
center.  The average distance from the Sun of the whole population of
dSphs is roughly 100~kpc.  Although these distances are small
compared to the size of the Local Group, they are
sufficiently large to have made proper motion measurements very
difficult using images obtained with ground-based telescopes.  The
effect of large distance on the size of the proper motion can be
lessened by a long interval between image epochs --- the time
baseline.  However, many of the nine dSphs were discovered only in the
past several decades (the most recent, Sagittarius, in 1994) and, thus,
the available images of dSphs with a quality high enough for proper
motion measurements have insufficient time baselines.

Despite these difficulties, several groups have reported proper motions
for a few dSphs.  Scholtz~\&~Irwin (1993) used Schmidt plates with a
time baseline of about 35~years to measure a proper motion,
$(\mu_{\alpha},\mu_{\delta})$\footnote{Throughout this article,
$\mu_{\alpha}$ is the proper motion in arcseconds in the direction of
increasing right ascension, \textit{i.e.}, the actual shift on the sky
in that direction.  The same is true for $\mu_{\ell}$, the proper
motion in the direction of increasing galactic longitude.}, for Draco
of $(90\pm50,100\pm50)$ milliarcseconds per century (mas~cent$^{-1}$,
hereafter) and a proper motion for Ursa Minor of
$(100\pm80,100\pm80)$~mas~cent$^{-1}$.  Schweitzer~\etal (1995) used a
variety of larger-scale plates with a time baseline of about 50~years
to measure a proper motion for Sculptor of
$(36\pm22,43\pm25)$~mas~cent$^{-1}$.  Similarly, Schweitzer \etal
(1997) used Palomar 5~m and KPNO 4~m plates with a time baseline of
42~years to measure a proper motion for Ursa Minor of
$(58\pm8,26\pm10)$~mas~cent$^{-1}$.  Finally, Irwin \etal\ (1996) used
Schmidt plates to measure a preliminary proper motion for Sagittarius of
210$\pm$70~mas~cent$^{-1}$.  All of the above proper motions are
corrected for the motion of the Local Standard of Rest (LSR, hereafter)
and the peculiar motion of the Sun with respect to the LSR.  The small
number and, in most cases, the large uncertainties of the existing
measurements emphasizes the difficulties inherent in measuring the proper
motion of a dSph.  Accurate proper motions of dSphs would explain or help to
answer a number of outstanding questions about the dSphs themselves,
the Milky Way, interactions between dSphs and the Milky Way, and the
formation of galaxies.  Probably the most fundamental question is
whether a dSph is gravitationally bound to the Milky Way.
Satellite status is almost universally assumed for the known dSphs ---
confirming this assumption requires, among other things, a
knowledge of the proper motion of each dSph.

	Using the Hubble Space Telescope (HST, hereafter) for
astrometry has advantages over using a ground-based telescope.  HST has
superior angular resolution which, in principle, allows a more precise
measurement of the location of an object.  This in turn reduces the
required time baseline between images of a dSph to just a few years.
As an example, consider an ``average'' dSph at a heliocentric distance
of 100~kpc with an assumed tangential velocity with respect to the Sun
of 220~km~s$^{-1}$.  Then the expected heliocentric proper motion of
the dSph is $46$~mas~cent$^{-1}$.  Assuming a plate scale of
51~mas~pixel$^{-1}$ for the CCD imager in HST cameras, the expected
shift in the position of the dSph over a one-year period is about
$0.009$~pixel.

Several ground-based astrometric programs (Monet \etal\ 1992; Tinney
\etal\ 1995; Tinney 1996) have shown that shifts of this magnitude can
be measured with CCD detectors.  CCD detectors used in differential,
small-angle astrometric measurements have significantly smaller
systematic errors compared to measurements with a long time baseline
using photographic plates, which are almost always acquired on
different telescopes and which have increased uncertainties because of
digitization on scanning machines.

	A single HST image of a stellar field is undersampled because
the full width at half of maximum (FWHM, hereafter) of a stellar
point-spread function (PSF, hereafter) is only about 1.0 pixel for the
HST imaging detectors.  Undersampling causes an image to have a
``digitized'' appearance and it makes measuring the centroid of a star
to an accuracy of few thousandths of a pixel impossible with just a
single image of the star.  Lauer (1999) and Anderson \& King (2000)
discuss these and other problems associated with undersampling and
offer remedies for these shortcomings.  They show that accurate
centroids of undersampled stellar images can be measured from multiple
images of a science field which are dithered in a pattern which, for
example, is an $N \times N$ grid of $1/N$ subpixel steps (Lauer 1999),
where $N^2$ is the number of images.  Such dithered images allow the
construction of a well-sampled effective point-spread function (ePSF,
hereafter), which is the convolution of the PSF of the telescope and
the function representing the spatial response of a pixel in the CCD
detector.  The value of the ePSF at a specific distance and direction
from its center is the response (formally the fluence: photon counts
per readout, though this article uses the more colloquial term flux)
measured by a pixel at that distance from the centroid of the star.
The centroid of an object and its brightness can be determined by
fitting the ePSF to the measured flux in an array of pixels
corresponding to the object.  Even though the individual functions in
the convolution defining the ePSF can be difficult, or even impossible,
to ascertain, the ePSF can be constructed without knowing an explicit
form for these functions.  Anderson \& King (2000) describe in detail a
method for constructing the ePSF and a procedure for fitting the ePSF
to the data in order to determine the centroid and total flux of an
object.  Bernstein (2002) provides additional discussion of these
topics.  Our method of constructing the ePSF, discussed in this
article, is very similar to that described in Anderson \& King (2000).

The expected small size of the proper motion of a dSph requires that
the reference frame be defined by extragalactic objects.  The images of
quasi-stellar objects (QSOs) are more compact at a given brightness
than are the cores of ordinary galaxies, making a QSO a better
reference frame than such a core.  For this reason (and because bright
QSOs are useful probes for gas in dSphs), Cudworth, Olszewski, \&
Schommer (1986), Tinney, Da Costa, \& Zinnecker (1997), and Tinney
(1999) (and references therein) undertook searches for QSOs behind the
dSphs.  Of the resulting QSOs, those that provide the required inertial
reference frame with the minimum exposure time have magnitudes
comparable to those of the brightest stars of the dSph that are in a
typical HST field.

	We are carrying out a program to measure proper motions for
four dSphs --- Carina, Fornax, Sculptor, and Ursa Minor --- from HST
images taken at three epochs, each separated by 1 -- 2 years.  This
article serves the following purposes.  1) It describes our method for
deriving an ePSF and for fitting it to the data to find the centroids
of objects.  2) It examines how accurately proper motions of the dSphs
can be measured using our methods and data.  3) Finally, the article
presents a preliminary proper motion of Fornax derived from two- or
three-epoch data for three distinct fields and discusses its
implications.

\section{Data}

	The data consist of images of fields in the Carina, Fornax,
Sculptor, and Ursa Minor dSphs obtained using HST with either the
Planetary Camera (PC; the high-resolution camera of the Wide Field
and Planetary Camera 2 -- WFPC2) with the F606W filter or the Space Telescope
Imaging Spectrograph (STIS) with no filter (50CCD).  Columns 1 and 2 in
Table~1 give the name of each field and the detector with which it was
imaged.  The third column gives the dates of the observations, both
existing and planned.  The fourth column gives the number of exposures
in the field and the average exposure time for each.  The remaining columns
give the right ascension, declination, galactic longitude, and galactic
latitude of the center of each field, respectively.  The fields are
approximately centered on known QSOs located in the
directions of the dSphs (Cudworth, Olszewski, \& Schommer 1986; Tinney,
Da Costa \& Zinnecker 1997; Tinney 1999).
The QSO in the FOR~$0240-3434$ field
is a gravitational lens with two images (Tinney 1995).  We currently
have three epochs of data for four fields and two epochs of data for
all but one of the remaining fields, where consecutive epochs are about
one or two years apart (see column 3 of Table~1).  At each epoch, there
are observations at eight dither positions.  The target dither pattern
consists of two quartets of pointings, where each quartet forms a
square with a side of 0.5~pixel.  The two quartets are shifted with
respect to each other by 0.25~pixel in both the $X$ and $Y$
directions.  This pattern forms a square grid with a spacing of
$\sqrt{2}$\,(0.25)~pixel rotated by an angle of 45~degrees with respect
to the $X$ axis.  For the STIS data, an error in setting up the dither
distorted the target pattern so that alternate lines of the square grid
are shifted along the line by $\sqrt{2}$\,(0.125)~pixel.  The error
appears to have had no effect on our ability to measure proper
motions.  The target pattern also includes integer pixel offsets to
shift the position of objects with respect to hot pixels.  The STIS
data has three images (cosmic ray splits) per dither position, while
for the PC data the number of images per dither position varies
between one and five.  The exposure time of the individual images is
the longest possible time that does not saturate the image of the QSO
and that satisfies the scheduling constraints imposed by the window of
visibility during the HST orbit.

The PC detector is a $800\times 800$~pixel CCD chip with a plate scale
of about 0.046~arcsec~pixel$^{-1}$.  Because of vignetting by the
pyramidal mirror, the usable area of a PC image is 34.4~arcsec $\times$
33.7~arcsec.  The STIS detector is a single $1024\times1024$~pixel CCD
chip with a plate scale of about 0.051~arcsec~pixel$^{-1}$.  Because of
vignetting near the edges of the detector, the usable area is only
approximately the central $50\times50$~arcsec$^2$.  The gain for all of
the PC data is 7 electrons per digital unit and for all of the STIS
data it is 1 electron per digital unit.

	The standard HST pipeline corrected the STIS and PC images for
the electronic bias, the dark current from hot pixels, and non-uniform
sensitivity (the flat-field).  In the case of the STIS images, the
pipeline also removes cosmic rays by combining the three images taken
at each dither position.  As part of this last correction, the pipeline
produces a record of the pixels affected by cosmic rays for both the
individual and the combined images.  The STIS pipeline produces three
arrays for each image (and each combined image): 1) the corrected pixel
photon counts, which we refer to as the science image or the science
data depending on the context; 2) the uncertainties in the pixel photon
counts; and 3) a word of 16 flags indicating (if non-zero) some problem
associated with the pixel (see Brown \etal\ 2002 for the definitions of
the flags).  The PC pipeline produces two arrays for each image: the
pixel photon counts and an array of flags (see Baggett \etal\ 2002).
The uncertainty of each pixel value for the PC images is calculated
from the value, gain, read noise, and a 0.3\% flat-field noise.

\section{Measurement of the Proper Motion}
\label{overview}

The task of measuring the proper motion of a dSph consists of four
broad steps:  1) identifying objects, 2) constructing the ePSF, 3)
determining centroids of objects by fitting the ePSF, and 4)
calculating the proper motion from the sets of centroids at different
epochs.  Steps 2) and 3) are inter-dependent because the construction of
the ePSF requires accurate centroids.  The remainder of this section
presents a brief introduction to these steps while a more detailed
discussion follows in the rest of the article.

Using the DAOPHOT package, we find the QSO and stars in the field and
determine their approximate location and brightness.  Our fields
contain relatively few stars, 100 -- 600, and most of them are faint.
This number of stars is a small fraction of those available in similar
published astrometric studies using HST (King \etal\ 1998; Bedin
\etal\ 2001; Kuijken \& Rich 2002).  The small number of stars in our
fields has an adverse impact on steps 2), 3), and 4).  For example, the
small number of stars limits both the accuracy of the ePSF and the
knowledge of changes in the ePSF with position within a field.  To
ameliorate the impact of the small number of stars on the accuracy of
the ePSF, we use all of the dither images available for a field at a
given epoch to generate a single ePSF.  This procedure effectively
increases the number of stars contributing to the ePSF, at the cost of
assuming that the ePSF neither varies among the images at one epoch
or within the images.

	If the image of the QSO is more extended than that of a star,
then including the QSO in the construction of the ePSF may distort the
ePSF.  Fitting such an ePSF could lead to systematic errors in the
centroids of both the stars and the QSO.  These systematic errors will
affect the measured proper motion if they are different at different
epochs.  The errors could differ among epochs if, for example, the
shapes of stellar images change because of guiding jitter.  The
$\chi^2$ resulting from fitting the ePSF to the QSO is similar to those
resulting from fitting to bright stars.  Thus, our procedure is to
include the QSO in the construction of the ePSF to improve the
signal-to-noise ratio ($S/N$) of the ePSF.

A quartic kernel provides the weight with which an observed pixel value
contributes to each point of the grid which defines the ePSF and the
average of these values is the initial estimate of the ePSF.
Iteratively applying the same interpolation and averaging scheme to the
difference between the observed pixel values and the ePSF refines the
estimate of the ePSF.  Forcing weighted moments of the ePSF to be zero
maintains a consistent centering of the ePSF on its grid among epochs.
The centroids of objects come from fitting the ePSF to observed pixel
values using least squares.  The estimation and fitting procedures are
similar for the PC and STIS images, with differences arising primarily
from the availability of combined images with cosmic ray rejection for
the STIS data.

The output of constructing and fitting the ePSF is a centroid for each
star in the field, $(X_{0,i}(t),Y_{0,i}(t))$ -- here $i$ is the
index of a star, and for the QSO, $(X_{0,\rm Q}(t),Y_{0,\rm Q}(t))$ at
at least two epochs:  $t_1$, $t_2$, \dots.  Also output is the
uncertainty in these centroids, which are calculated from the scatter of
the individual values derived from each image at a given epoch around
their mean.

The proper motion of the dSph causes a change in the average position
of its stars with respect to the QSO.  The measured centroids differ
among epochs both because of the proper motion and because of
differences in pointing, image rotation, and image scale.  Measuring
the proper motion requires a consistent coordinate system among all
epochs.  But the centroid of the QSO at two epochs only contains enough
information to determine the difference in pointing.  Thus, we use the
stars of the dSph to define a standard coordinate system and
determine the four-parameter transformation ($X$ and $Y$ offsets,
rotation, and relative scale) that relates the coordinate systems at
different epochs.  The measured proper motion of the dSph is then the
opposite of the change in the centroid of the QSO in the standard
coordinate system.

Geometrical distortions introduced by the STIS and WFPC2 optics could
require a more complex transformation between the coordinate systems at
pairs of epochs.  A more complex transformation has more degrees of
freedom and, therefore, requires more stars to determine it
accurately.  Given the small number of stars in a typical field,
introducing more than four parameters into the transformation could
degrade significantly the accuracy of the measured proper motion.  We
minimize the need for a more complex transformation by both removing
the best estimates of the distortions and, more importantly, by putting
objects back on the same pixel of the detector as closely as possible
at the different epochs.  The latter ensures that any uncorrected
distortion is the same at both epochs and, so, does not affect the
measured proper motion.

	The two components of the change in the centroid of an object
in the standard coordinate system between two epochs are
\begin{eqnarray}
p_{x} &=& X_{0}^{tr}(t_2) - X_{0}(t_1)
\label{px} \\
p_{y} &=& Y_{0}^{tr}(t_2) - Y_{0}(t_1).
\label{py}
\end{eqnarray}
In the above equations, $(X_{0}^{tr}(t_2), Y_{0}^{tr}(t_2))$ are the
coordinates of an object at the second epoch transformed into the standard
coordinate system of the first epoch, $t_1$.  The uncertainties in $p_{x}$
and $p_{y}$ have contributions from the uncertainties of the measured
centroids at the two epochs and from the uncertainty in the transformation.
The proper motion of the dSph is proportional to $-p_{x}$ and $-p_{y}$
of the QSO.

\section{Selecting Objects}
\label{preliminaries}
\subsection{Initial Estimate of Centroids for Objects}

	The initial estimate of the centroids of objects and of their
magnitudes comes from processing the images with the DAOPHOT and
ALLSTAR software packages (Stetson 1987, 1992, 1994).  The DAOPHOT task
FIND distinguishes objects in the images.  The objects are stars (stars
of the dSph and foreground stars of our Galaxy), galaxies, QSOs, and,
unwontedly, spurious objects such as cosmic rays and ``hot pixels.''  A
visual examination of the images eliminates galaxies and close pairs
of stars from further
consideration.  Photometry based on PSF fitting by ALLSTAR gives the
locations and magnitudes of objects.  Subsequently, DAOMATCH and
DAOMASTER match the objects found in individual images and derive
translations between the locations in different images.  The
master list of all objects found in the images and the set of 
translations provide the first estimate of the centroids of
objects.

\subsection{Extracting Objects from Images}

\begin{figure}[p]
\centering
\includegraphics[angle=-90,scale=0.8]{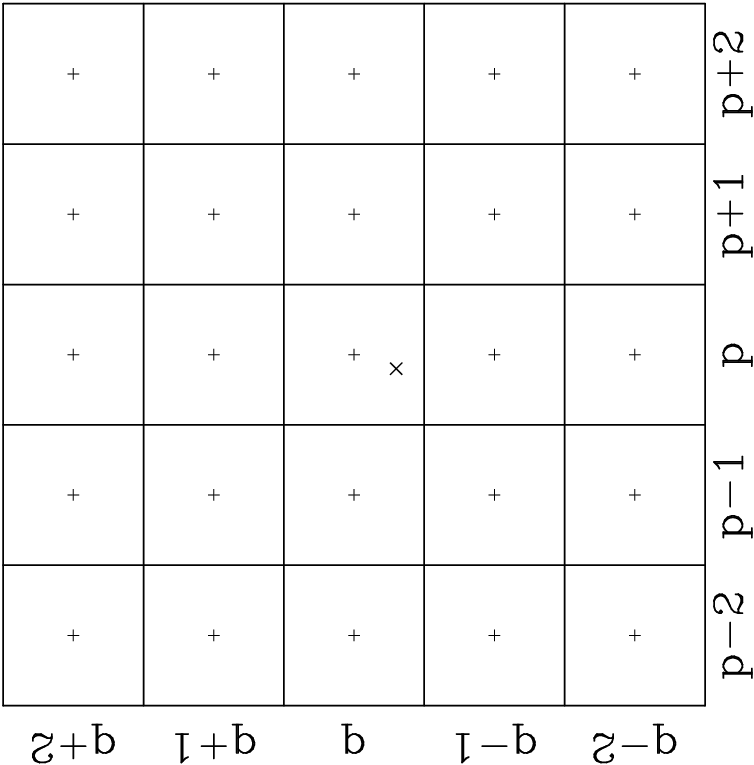}
\caption{The $5\times5$ array of science data pixels extracted for
each object.  A plus marks the center of a pixel and the slanted cross
marks the centroid of an object.}
\label{avarray}
\end{figure}

	Because of the undersampling in the PC and STIS CCDs, all of
the information about the brightness and location of an object is
contained in only a few pixels.  For convenience, we extract from the
science image a $5\times5$ array of pixel values centered on the
DAOPHOT centroid of an object as shown in Figure~\ref{avarray}.  Here
$p$ and $q$ are the indices of the pixel that contains the centroid of
the object, $(X_{0},Y_{0})$.  In the figure, a plus marks the center of
each pixel and the slanted cross marks the centroid of the object.
The relationships between $(X_{0},Y_{0})$ and $(p,q)$ are
\begin{eqnarray}
p &=& Int(X_{0}) + 1 \label{pav} \\
q &=& Int(Y_{0}) + 1. \label{qav}
\end{eqnarray}
In these equations, the function $Int$ takes the integer part of its
argument.  Our convention for indexing pixels and for the coordinates
of the center of a pixel determine Equations~\ref{pav}\ and \ref{qav}:
the pixel in the lower left corner of an image has indices $(1,1)$ and
its center has coordinates $(0.5,0.5)$.  Note that this convention for
the coordinates of the center of a pixel differs from that used in
DAOPHOT, where the center of the lower left pixel has coordinates
$(1.0,1.0)$.  Our computer code takes this difference into account when
it reads the master list.

	For each object we also extract from the FITS files the
corresponding $5\times5$ array containing data-quality flags and, for
STIS, the corresponding $5\times5$ array containing the uncertainties
of the pixel values.

\subsection{Sky Subtraction}

	The signal in the $5\times 5$ science data array comes from the
central object and from ``the sky'' --- scattered light and light from
unresolved galaxies and stars.  Our images are empty enough that an
accurate estimate of the sky is the median value of those pixels whose
centers are in an annulus centered on the object with inner and outer
radii of 8.0 and 16.0 pixels, respectively.  The sky is subtracted from
the signal of each pixel in the science data array to ensure that only
light from the central object contributes to the construction and
fitting of the ePSF.

The value of the sky in the PC images is very small: about one digital
unit.  Because of the wider bandpass of the unfiltered STIS images
and the smaller number of electrons per digital unit, the
value for the sky in these images is a few tens of digital units.  The
sky does not vary significantly with position in an image for either
the PC or STIS data.  The sky level does vary between images, which
probably reflects different amounts of scattered light at different
points in the HST orbit.  For the STIS data, the measured sky value
increases with increasing flux of an object when its total flux in the
science data array is larger than about $10^4$~digital units.  Above
this value of the total flux, the contribution from the ``tails'' of
the flux distribution of an object to the signal in the sky annulus
becomes larger than the noise in the estimate of the sky.  However,
this effect is negligible since the higher skies for these bright
objects are very small when compared to the total fluxes of these
objects.

\section{Construction of the ePSF}
\label{constructepsf}
\subsection{Selecting Objects for the Construction of the ePSF}

	A star, or QSO, which contributes to the construction of the
ePSF must satisfy two criteria.  1. Certain specified pixels in the
$5\times5$ science data array must have zero flags.  2. The $S/N$ of an
object must be larger than some specified limit.  Below we discuss each
criterion in greater detail.

There are generally very few objects in our images that are both bright
and that have all of their flags equal to zero.  Experience showed that
including objects that had a few pixels with non-zero flags did not
significantly degrade the final result.  The analysis in this article
requires that the flags be equal to zero for the inner $3\times 3$
pixels of the science data array.

	The second criterion requires that the $S/N$ of an object
calculated with
\begin{equation}
S/N=\frac{\sum_{i=1}^{25} v_{i}}
{\left(\sum_{i=1}^{25}u_{i}^2\right)^{1/2}}
\label{sigtonoise}
\end{equation}
be larger than some specified limit.  Here $v_{i}$ is the signal in a
pixel (corrected for the sky) and $u_{i}$ is the uncertainty in
$v_{i}$.  Both sums include only those pixels whose flags equal zero.

	Trade-offs determine the value of the $S/N$ limit.  A
relatively large $S/N$ limit allows only a few very bright objects to
contribute to the construction of the ePSF.  The small number of
contributing objects introduces noise into the ePSF due to the
undersampling.  Decreasing the $S/N$ limit allows more objects to
contribute, reducing the effects of undersampling, but progressively
increasing the effect of photon noise on the constructed ePSF.  The
adopted $S/N$ limit is the one which minimizes the uncertainties in the
fitted stellar centroids of the brightest $10\%$ of the stars.
However, the final results are insensitive to the exact value of this
limit.  For Fornax, the limit is $S/N = 30$.  In general, the adopted
value could vary from field to field depending on the number of stars
present.

	Our fields have relatively few stars.  For example, the master
list has about 600 entries for the field with the largest number of
stars and only about 100 entries for the field with the smallest.  Of
these, only a handful are bright stars.  Thus, the scarcity of bright
stars is a serious problem for the construction of the ePSF.  The best
sampling of the ePSF results from constructing a single ePSF per field
at each epoch.  For the STIS data, constructing an ePSF from either the
24 individual science images or the 8 combined images with cosmic ray
rejection yields statistically indistinguishable average uncertainties
of the centroids of the brightest 10\% of the stars.  The results in
this article are for 24 images.  For the PC data, the HST
pipeline processing does not produce combined images and, thus, the
results in the article are for all of the images.

\subsection{Construction of the ePSF: Initial Estimate}
\label{psfie}

	Our method constructs a single ePSF for all of the images of a
field at an epoch.  The ePSF is realized on a $25\times25$ grid with a
spacing between grid points of 1/5~pixel.  Bicubic spline
interpolation gives the value of the ePSF between grid points.  An
individual grid point has indices $\alpha$ (horizontal direction) and
$\beta$ (vertical direction).  The mathematical notation for the ePSF
is $\Pi_{\alpha,\beta}$ and the notation for its normalized form is
$\hat{\Pi}_{\alpha,\beta}$.

	Each object contributing to the construction of the ePSF must
satisfy the aforementioned $S/N$ criterion and have the flags for all
of the inner $3\times3$ array of pixels in the science data array equal
to zero. Each object contributes with the same weight, which is ensured
by normalizing the data arrays of the objects according to
\begin{equation}
\hat{v}_{i} = \frac{v_{i}}{\sum_{j=1}^{25} v_{j}} \label{normal}
\end{equation}
so that
\begin{equation}
\sum_{i=1}^{25} \hat{v}_{i} = 1.
\end{equation}
In the equations, $\hat{v}_{i}$ is a normalized science pixel
value with an index $i$ for a particular object.  The sums
include only those pixels that have flags equal to zero.

\begin{figure}[p]
\centering
\includegraphics[angle=-90,scale=0.8]{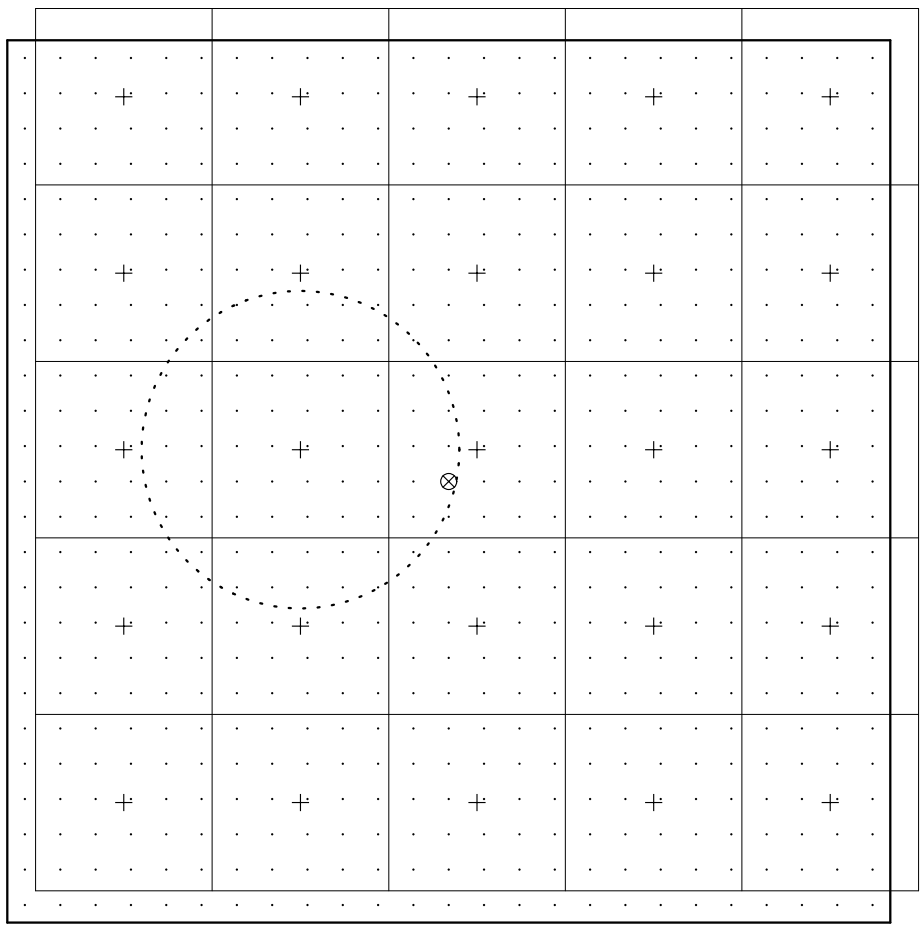}
\caption{The solid $5\times5$ grid represents the science data array
for a particular object.  A cross marks the center of each science
data pixel.  The slanted cross marks the centroid of the object.  The
$25\times25$ array of dots represents the points at which the ePSF is
realized.  The center of the ePSF grid, marked with an open circle,
coincides by definition with the centroid of the object.  Since the
centroid of an object can be anywhere within the central pixel in the
$5\times5$ grid, the two grids are offset by $\vec{\delta}$, whose
magnitude is equal to the distance between the centroid of the object
and the center of the central science data pixel.  The open dashed
circle encloses those ePSF grid points to which the science data pixel
on which the circle is centered contributes.  The radius of the dashed
circle is an adjustable parameter, $k$, referred to as the ePSF
smoothing length.}
\label{conpsf}
\end{figure}

\begin{figure}[p]
\centering
\includegraphics[angle=-90,scale=0.90]{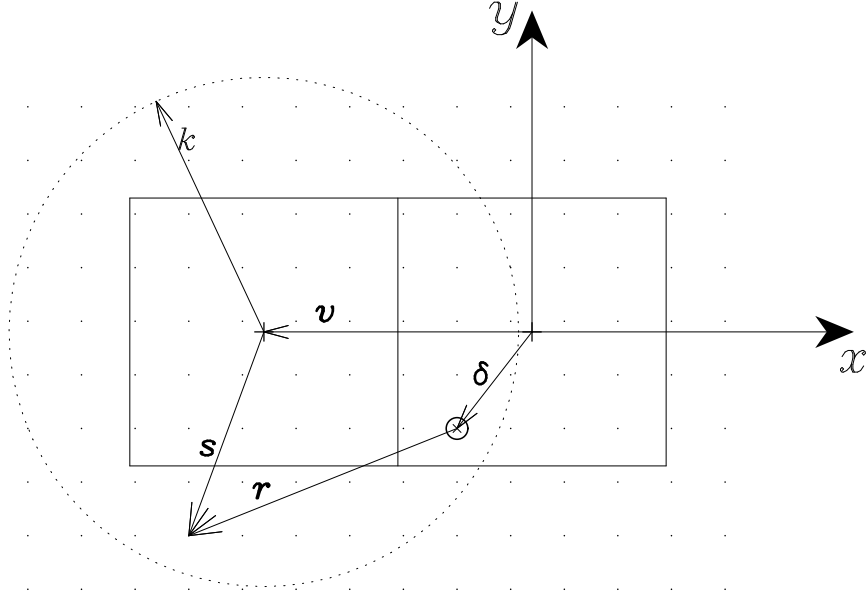}
\caption{The figure shows a relationship among the quantities involved
in the construction of the ePSF.  The two solid squares are science
data pixels: the central pixel, whose center defines the origin of a
local $x$-$y$ coordinate system, and an adjacent one which is
contributing to the ePSF grid point.  The pluses mark their centers.
The dots represent the ePSF grid points.  The center of the ePSF,
marked with an open circle, coincides with the centroid of the object,
marked with the slanted cross.  The science data pixel with the
position vector $\vec{v}$ contributes to the ePSF grid point with the
position vector $\vec{r}$.  The contribution is weighted, where the
weight depends on the length of vector $\vec{s}$ and is calculated from
Equation~\ref{weight}.  The science data pixel contributes to all of
the ePSF grid points contained in the dashed circle, whose radius is
equal to the adjustable smoothing length, $k$.  From the figure, it
follows that $\vec{\delta} +\vec{r}=\vec{v}+\vec{s}$.  Note that
$\vec{v}$ is not always horizontal and it should not be confused with
$v$, which denotes a science data pixel value.}
\label{conpsfdet}
\end{figure}

	Figure~\ref{conpsf} shows the relationship for a particular
object between the
$5\times5$ science data array (delineated with solid lines) and the
$25\times25$ ePSF grid, where dots mark the grid points.
A cross marks the center of each science data pixel and the slanted
cross marks the centroid of the object, $(X_{0},Y_{0})$.  The center of
the $25\times25$ ePSF grid, marked by the small open circle, coincides by
definition with $(X_{0},Y_{0})$ and, thus, with the slanted cross.  The
two grids are offset from each other by $\vec{\delta} = (\delta_x,
\delta_y)$. Equations~\ref{delx} and \ref{dely} give the components of
$\vec{\delta}$ and Figure~\ref{conpsfdet} shows its relationship to
other quantities discussed below.
\begin{eqnarray}
\delta_x&=& X_{0}-p+\frac{1}{2} \label{delx} \\
\delta_y&=& Y_{0}-q+\frac{1}{2} \label{dely}
\end{eqnarray}
An open dashed circle centered on the center of a science data pixel in
Figure~\ref{conpsf} encloses all of the ePSF grid points to which this
pixel contributes.  The radius of this dashed circle is an
adjustable ePSF smoothing length, $k$.  The contribution that a
science data pixel with index $i$ makes to the ePSF at a grid point
with indices $(\alpha,\beta)$ has a weight given by
\begin{equation}
w_{i,\alpha,\beta} = \left\{ \begin{array}{ll}
  \left(1-\frac{\vert \vec{s} \vert ^{2}}{k^{2}}\right)^{2} &
     \mbox{if $\vert \vec{s} \vert < k$}  \\
  0 & \mbox{if $\vert \vec{s} \vert > k$}.
\end{array} \right.
\label{weight} 
\end{equation}
In the above, $\vert \vec{s} \vert$ is the distance between
the center of the science data pixel and the ePSF grid point and is given by
\begin{equation}
\vert \vec{s} \vert = \vert \vec{r} + \vec{\delta} - \vec{v} \vert .
\label{sep}
\end{equation}
Here, $\vec{r}$ is the vector from the central ePSF grid point to the ePSF
grid point $(\alpha,\beta)$ and $\vec{v}$ is the vector from the center of
the central science data pixel to the contributing science data pixel.
See Figure~\ref{conpsfdet}, which shows the relationship among the
quantities just described.

	The initial estimate of the ePSF at each grid point $(\alpha, \beta)$
is the weighted average of the normalized values of the contributing pixels
of the science data arrays:
\begin{equation}
\Pi_{\alpha,\beta} = \left(\sum_{\rm images} \sum_{\rm objects} \sum_{i=1}^{25}
\hat{v}_{i} w_{i,\alpha,\beta}\right) /
\left(\sum_{\rm images} \sum_{\rm objects} \sum_{i=1}^{25}
w_{i,\alpha,\beta}\right).
\label{psf}
\end{equation}
The normalized ePSF is
\begin{equation}
\hat{\Pi}_{\alpha,\beta}=\frac{25\, \Pi_{\alpha,\beta}}
{\sum_{\mu=1}^{25}\sum_{\nu=1}^{25} \Pi_{\mu,\nu}},
\label{psfnone}
\end{equation}
where the factor of $25$ is the number of ePSF grid points per
science data pixel.

At this stage, the ePSF does not have the optimal shape because: 1) the
centroids of objects contributing to its construction are only initial
estimates; 2) the smoothing controlled by the parameter $k$ broadens
the ePSF; and 3) the finite size of the $5\times 5$ science data array
causes the kernel estimate of the ePSF to be too high at grid points
that are closer than a smoothing length from the boundary of the grid.
In addition, the summation in Equation~\ref{psf} does not guarantee
that the ePSF is centered properly.   The following section discusses
the issue of centering and the section after that returns to the first
three points.

\subsection{Construction of the ePSF: Centering}

	The center of the ePSF grid defines the centroid of an object
fitted to the ePSF (see Figure~\ref{conpsf}).  Thus, the ePSF must be
centered on its grid in a repeatable way at all epochs.  The
construction of the ePSF described in the previous section does not
enforce any centering of the ePSF.  A simple way to center the ePSF is
to shift the ePSF so that its largest value is at the center of the
grid.  However, this is not the best way to center the ePSF because the
slow variation of the ePSF near its peak makes the location of the
maximum particularly susceptible to noise.  An alternative way is to
place the ``center of light'' of the ePSF at the center of the grid.  A
weakness of this approach is that the position of the ``center of light'' is
particularly sensitive to values of the ePSF far from the center.
These values provide little information on the centroids of objects and
are also affected by noise and by errors in the level of the
subtracted sky.  The effect of the distant grid points can be reduced
by calculating a ``center of light'' with the largest weight for those grid
values that contain the greatest amount of information about the
centroid.

	In simplest terms, the determination of the centroid of an
object involves shifting the ePSF with respect to the science data
array to minimize some quantitative measure of the difference between
the values of the ePSF and of the science data pixels.  Most of the
information about the centroid of the object comes from the region
where the ePSF varies most rapidly, since this is where a small offset
between the ePSF and the object produces the largest difference.  This
region is between the relatively flat inner and outer sections of the
ePSF.  These ideas argue that the grid points where the ePSF varies
most rapidly should have the greatest influence on the location of the
center of the ePSF.  Thus, our algorithm centers the ePSF by forcing a
weighted ``center of light'' to coincide with the central grid point.  The
weights are proportional to the derivative of a Gaussian function,
\begin{equation}
\label{gweight}
w(\vec{r})=\frac{\vert \vec{r} \vert}{r_{0}} \exp\left (-\frac{\vert \vec{r}
\vert ^2}{2 r_{0}^2} \right ).
\end{equation}
Here $\vert \vec{r} \vert$ is the distance from the center of the ePSF
grid and $r_{0}$ is an adjustable parameter on the order of 0.5 $\times$
FWHM of the ePSF.  Then the weighted ``center-of-light'' of the
ePSF, $\vec{r}_{\rm cm}$, is
\begin{equation}
\label{cmpsf}
\vec{r}_{\rm cm}=\frac{\sum_{\alpha=1}^{25}\sum_{\beta=1}^{25} 
\vec{r}_{\alpha,\beta}
w(\vec{r}_{\alpha,\beta}) \hat{\Pi}(\vec{r}_{\alpha,\beta})}
{\sum_{\alpha=1}^{25}\sum_{\beta=1}^{25}
w(\vec{r}_{\alpha,\beta}) \hat{\Pi}(\vec{r}_{\alpha,\beta})}.
\end{equation}
The algorithm iteratively shifts the ePSF on the grid by $-\vec{r}_{\rm cm}$
using bicubic spline interpolation,
\begin{equation}
\Pi(\vec{r}_{\alpha,\beta}+\vec{r}_{\rm cm}) \rightarrow
\Pi(\vec{r}_{\alpha,\beta}),
\end{equation}
until the $x$ and $y$ components of $\vec{r}_{\rm cm}$ are
smaller than 0.001~pixel.  Here and for the rest of this discussion we
drop the ``hat'' notation, since by the ePSF we will now always mean
its normalized form.

\subsection{Construction of the ePSF: Iterative Adjustments}
\label{iterative}

	The initial estimate of the ePSF differs from the true ePSF for
the reasons described at the end of Section~\ref{psfie}.  Our iterative
algorithm for improving the shape of the ePSF begins by deriving new,
presumably more accurate, centroids by fitting the initial estimate of
the ePSF to the science data arrays.  An improved shape of the ePSF
then results from calculating the difference between each pixel of the
science data array of an object and the fitted ePSF and then
calculating the weighted average of this difference at each ePSF grid
point using all of the objects contributing to the ePSF.  Adding this
correction to the original ePSF produces a more accurate estimate.  The
weights are again calculated with equation~\ref{weight}.  The
difference between the fitted ePSF and the data varies more slowly than
the ePSF itself and the shallower gradients reduce the bias near the
boundary of the ePSF grid --- the bias caused by smoothing with a
kernel.

The equation that implements these corrections is
\begin{equation}
\Pi_{\alpha,\beta} \leftarrow \Pi_{\alpha,\beta}+
\left(\sum_{\rm images} \sum_{\rm objects} \sum_{i=1}^{25}
\frac{v_{i}^{j}-f^{j}\Pi(\vec{v}_{i}^{j}-\vec{\delta}^{j})}{f^{j}}
w_{i,\alpha,\beta}\right) / \left(\sum_{\rm images} \sum_{\rm objects}
\sum_{i=1}^{25} w_{i,\alpha,\beta}\right).
\label{corrpsf}
\end{equation}
In the above equation, $f^{j}$ is the total fitted flux of object $j$.
The vector $\vec{v}-\vec{\delta}$ contains the implicit dependence of
the correction on the centroid of an object.  The iterative application
of Equation~\ref{corrpsf} gives progressively more accurate centroids
and a progressively more accurate estimate of the ePSF.  Of course,
noise in the science data eventually limits the usefulness of
subsequent iterations.   We find that the ePSF converges to a stable
solution in four iterations.

The construction of the ePSF for PC data has an additional step that
identifies and excludes those science data pixels which are corrupted
by cosmic rays or large dark rates.  The initial step is to calculate
the median for the
pixels in each particular location in the science data arrays of
an object and to exclude those values which are more than five times the
median.  After one iteration of corrections has improved the
ePSF, fitting the ePSF to those science data arrays contributing to the
construction of the ePSF yields the expected value of each pixel and
those pixels which are more than five times their uncertainties from
this expected value are excluded from further consideration.  The
uncertainty includes both the uncertainty in the pixel value and
6\% of that value, where the latter accounts for the mismatch between
the constructed and true ePSF.  Using the
remaining pixels, the algorithm updates the list of objects
contributing to the ePSF and begins the construction the ePSF anew.
This process of identifying cosmic rays and constructing the ePSF anew
iterates until the number of additional corrupted pixels is zero.
The algorithm then continues with the iterative corrections to the
ePSF as for the STIS data.  In the STIS data, a few pixels with large
dark rates (``hot pixels'') may not be flagged by the reduction pipeline
but could be eliminated by the methods described in this paragraph.
We have not found this to be necessary for our Fornax images, but we will
consider it on a case-by-case basis for our other dSph data.

	An additional refinement to the centroid of an object in the
iterative loop applying corrections to the ePSF is to replace it with
the average calculated from the centroids of the object in different
images.  In detail, the procedure is as follows.  Find all of the
objects common to the fiducial image (chosen to be the first image) and
an image k.  Let $\vec{X}_{i}^{1}$ and $\vec{X}_{i}^{k}$ be the
position vectors of the centroids of the same object, ${i}$, in the
fiducial and $k$ images, respectively.  Note that generally
$\vec{X}_{i}^{1} \neq \vec{X}_{i}^{k}$ because of dithering and a
possible difference in image scale and orientation.  Determine the
ratio of the image scales, $s$, offset, $\vec{D}$, and angle of
rotation, $\theta$, in the transformation
\begin{equation}
\vec{X}^{1}=s^{1,k}{R}(\theta)^{1,k}\times(\vec{X}^{k}-\vec{D}^{1,k}).
\label{tran}
\end{equation}
with a least-squares fit.  Here $R(\theta)$ is the solid-body
rotation matrix:
\begin{equation}
R(\theta)=\left (\begin{array}{rr} \cos(\theta) &\sin(\theta) \\
-\sin(\theta)&\cos(\theta) \end{array} \right ).
\label{rotmatrix}
\end{equation}
Transform the centroid of an object in image $k$ to the coordinate
system of the fiducial image.  Find the average centroid using the
value from every image where the object satisfies the criteria for
contributing to the construction of the ePSF.  Transform the average
centroid back to the coordinate system of image $k$.

\begin{figure}[htp]
\hspace{-0.6truein}
\includegraphics[angle=-90,scale=0.90]{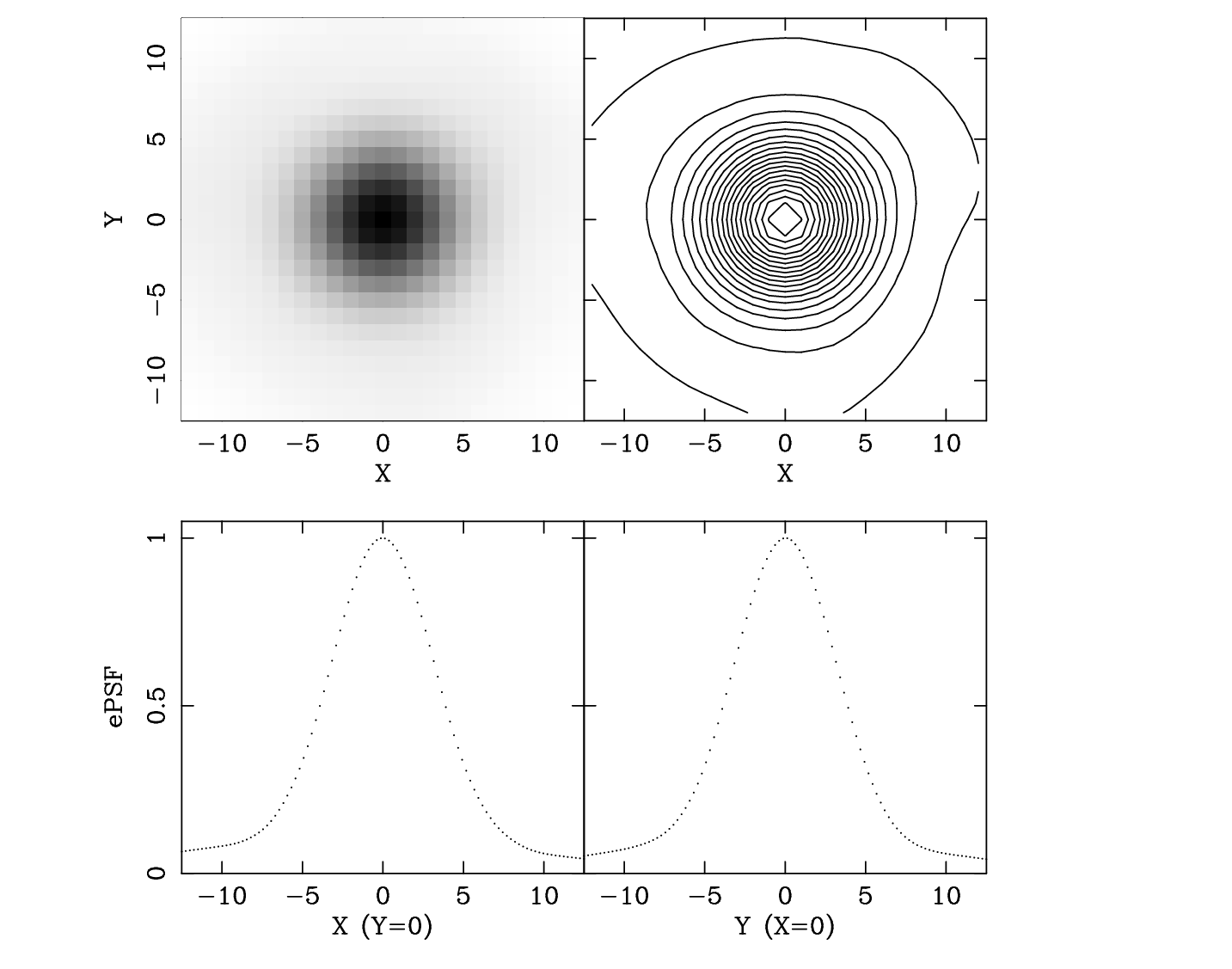}
\caption{The ePSF after four iterations for the 24 first-epoch STIS
images of the FOR~J$0238-3443$ field.  Top-left panel: Gray-scale map
of the ePSF.  Top-right panel: Contour plot of the ePSF; the levels
are separated by 5\% of the maximum value.  Bottom-left panel: ePSF
values along the horizontal bisector.  Bottom-right panel: ePSF values
along the vertical bisector.}
\label{plotpsf}
\end{figure}

	The panels in Figure~\ref{plotpsf} show the ePSF resulting
after four iterations for the 24 first-epoch STIS images of the
FOR~J$0238-3443$ field as a gray-scale map (top-left) and a contour plot
(top-right) and also show its cross section along the horizontal
bisector (bottom-left) and the vertical bisector (bottom-right).  The
levels in the contour plot are separated by 5\% of the maximum value.
A total of 738 science data arrays from the 24 images contributed to
the construction of the ePSF (note that a given object contributes at
least one and at most 24 science data arrays).  Each occurrence of an
object must have a $S/N$ larger than 30 and the flags of the inner
$3\times3$ of the $5\times5$ array of science data pixels equal to 0.
The smoothing length, $k$, is 1~pixel and the width of the weighting
function used in the centering algorithm, $r_{0}$, is 0.8~pixel.  Based
on experience, these values of the free parameters are close to the
optimal values which yield the smallest uncertainties in the positions
of objects.

\begin{figure}[htp]
\hspace{-0.6truein}
\includegraphics[angle=-90,scale=0.90]{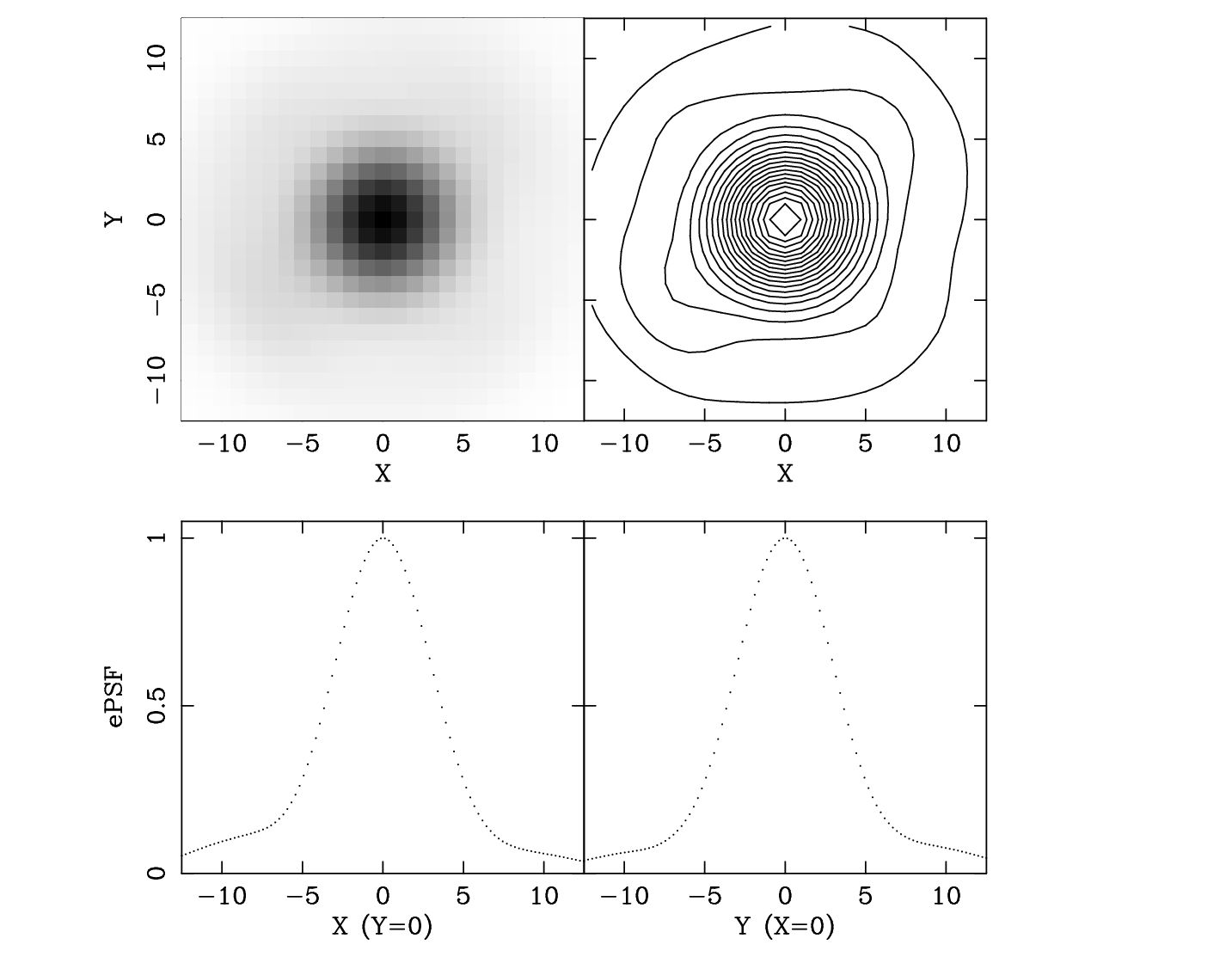}
\caption{The ePSF after four iterations for the 18 first-epoch PC
images of the FOR~J$0240-3434$ field.  Top-left panel: Gray-scale map
of the ePSF.  Top-right panel: Contour plot of the ePSF; the levels
are separated by 5\% of the maximum value.  Bottom-left panel: ePSF
values along the horizontal bisector.  Bottom-right panel: ePSF values
along the vertical bisector.}
\label{plotpsfpc}
\end{figure}

Similarly, Figure~\ref{plotpsfpc} shows the ePSF resulting after four
iterations for the 18 first-epoch PC images of the FOR~J$0240-3434$
field.  The panels are the same as Figure~\ref{plotpsf}.  A total of
513 science data arrays from the 18 images contributed to the
construction of the ePSF, with each having a $S/N$ larger than 30 and
the same requirements for the flags as for Figure~\ref{plotpsf}.  The
values for $k$ and $r_{0}$ are 1.0~pixel and 0.8~pixel, respectively.

	The circular and concentric contours in the inner region of the
ePSF, as depicted in the top-right panels in Figures~\ref{plotpsf} and
\ref{plotpsfpc}, show that the ePSF is azimuthally symmetric in this
region.  The outer $3 - 4$ contours, for which the values of the ePSF
are smaller and noisier than those in the inner region, deviate from
azimuthal symmetry.  The outer part of the ePSF for the PC data has
extensions pointing towards the four corners of the figure that are
probably caused by diffraction spikes. The FWHM of the ePSF is about
1.6~pixel (or about 8 ePSF grid spacings) for the STIS data and about
1.4~pixel for the PC data.  The value of the FWHM depends on the choice
of the ePSF smoothing length, $k$.  It increases with increasing $k$.

	There is no way to tell from Figures~\ref{plotpsf} and
\ref{plotpsfpc} whether the depicted ePSF is the true ePSF.  A way to
test the algorithm that constructs and centers the ePSF is to apply it
to artificial images that closely model real data.  The artificial data
consist of 8 images where the dithers are the same as that for the
FOR~J$0238-3443$ field.  The artificial field contains 962 stars located
on a grid with centroids at random locations within a pixel.  The
values of the science data pixels are generated with a two-dimensional
Gaussian distribution with a FWHM equal to 1.5 pixels in both
dimensions.  The noise in the images has a normal distribution and an
amplitude given by the uncertainty of a pixel value in the science data
array.  Furthermore, we add ``sky'' to the images and set the flags of
$3\%$ of the pixels to a non-zero value. The latter reflects the
percentage of non-zero flags in the real data.

\begin{figure}[hp]
\includegraphics[angle=-90,scale=0.7]{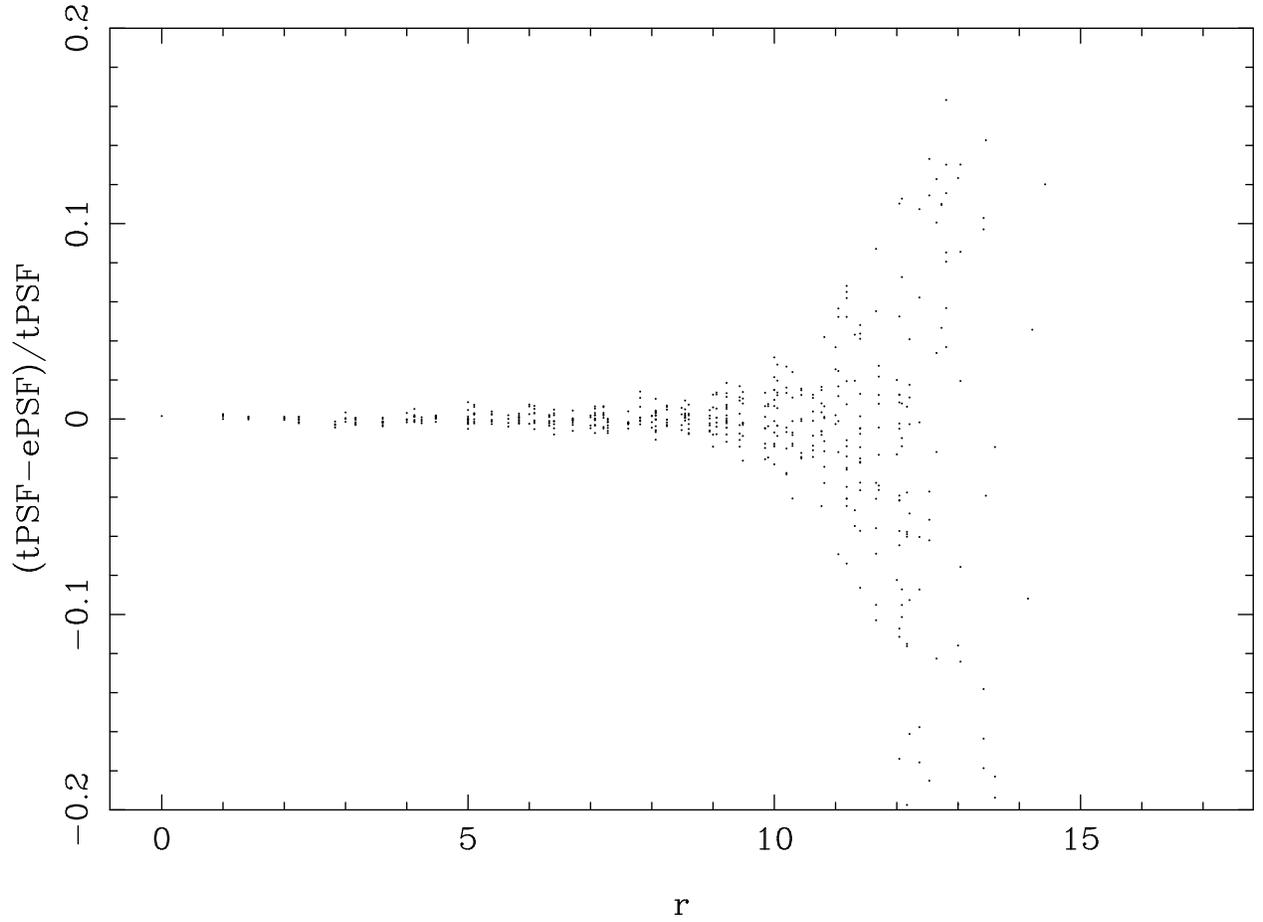}
\caption{The fractional difference between the true PSF (tPSF) and the
constructed ePSF, after four iterations, as a function of the distance
from the center of the ePSF grid.  The tPSF is a two-dimensional
Gaussian function with a FWHM of 1.5~pixels in both directions.  Note
that 1.0~pixel is five times the spacing of the ePSF grid --- the grid
spacing is the unit of distance in the plot.}
\label{tpsfepsf}
\end{figure}

	Figure~\ref{tpsfepsf} shows the fractional difference between
the true PSF (tPSF) and the constructed ePSF for the artificial
images, after four iterations, as a function of the distance $\vert
\vec{r} \vert$ from the center of the ePSF grid.  A total of 5843
science data arrays contributed to the construction of the ePSF. The
smoothing length, $k$, is 0.5~pixel (2.5 in the units of the plot)
and the width of the weighting function in the
centering algorithm, $r_{0}$, is 0.8~pixel.  The mean fractional
difference between the tPSF and the ePSF is less than 0.5\% for
grid points with $\vert \vec{r} \vert \la 10$ (or about 2.0 pixels).
The mean difference is small even at larger radii, but the increasing
scatter in the points makes this mean uncertain.  The larger scatter in
the points at larger radii is a reflection of the small values of the
tPSF there.  The scatter in Figure~\ref{tpsfepsf} reflects the noise in
the ePSF.  The noise increases as the smoothing length decreases below
0.5~pixel.  Increasing the smoothing length decreases the noise,
however it also introduces systematic differences between the ePSF and
tPSF as the smoothing length approaches the FWHM of a stellar image.
For example, the fractional difference shows systematic trends with
$\vert \vec{r} \vert$ that are 10\% or larger for $k = 1.3$~pixel and a
FWHM of 1.5~pixel.  Figure~\ref{tpsfepsf} shows that the algorithm which
generates the ePSF reproduces the tPSF if the smoothing length is not
too large compared to the stellar FWHM.

	Reducing the number of artificial stars contributing to the
construction of the ePSF from 5843 to 128 (the latter is of the same
order as the number of stars used to construct the ePSF for the real
data) increases the noise in the ePSF.  The noise increases both
because of the smaller number of photons contributing to the ePSF and
the sparser sampling of the tPSF.  For $k=0.5$, the mean fractional
difference between the tPSF and the ePSF is again about $0\%$ for all
values of $\vert \vec{r} \vert$.  However, the root-mean-square
(\textit{rms}) scatter is larger than that in Figure~\ref{tpsfepsf}.
It increases with increasing $\vert \vec{r} \vert$ and is about $1\% -
2\%$ for $\vert \vec{r} \vert \leq 5$.

	Selecting a value for the smoothing length, $k$, is a balance
between decreasing the noise and minimizing the systematic
errors in the ePSF.  Its optimal value minimizes the average
uncertainty in the positions of the brightest stars.  The uncertainty
in the position of a star is the uncertainty in the mean of the positions
from the individual images as estimated from their scatter.

\subsection{Is the ePSF for a QSO different from that for a star?}
\label{qpsfspsf}

	The objects contributing to the construction of the ePSF are
stars and the QSO.  Using the same ePSF to determine the centroids of
both the stars and the QSO could produce systematic errors in the
centroids if the ePSF of the QSO differs from the ePSF of a star.
These systematic errors cancel out if they are the same at all epochs.
The cancellation would occur if the ePSF is the same at all epochs, but
this may not be the case.  Indeed, our data show evidence that the ePSF
can differ between epochs:  the middle epoch of the FOR~J$0240-3438$
field has an ePSF that varies across the image, whereas the other two
epochs do not show such variability (see Section~\ref{psfdiag}).  Thus, it is
necessary to explore whether the ePSF of the QSO is the same as that of
a star.

Visual inspection of the images of the QSOs in our three fields in
Fornax show that all appear stellar, but a quantitative comparison
between the ePSF of a QSO and that of the stars might reveal more
subtle differences.  One such method is to construct both ePSFs and
compare them.  However, this method is not feasible for our data since
there are too few images of a QSO to construct a well-sampled ePSF for
it --- the noise introduced by the undersampling dominates the
comparison.

\begin{figure}[p]
\centering
\includegraphics[angle=-90,scale=0.85]{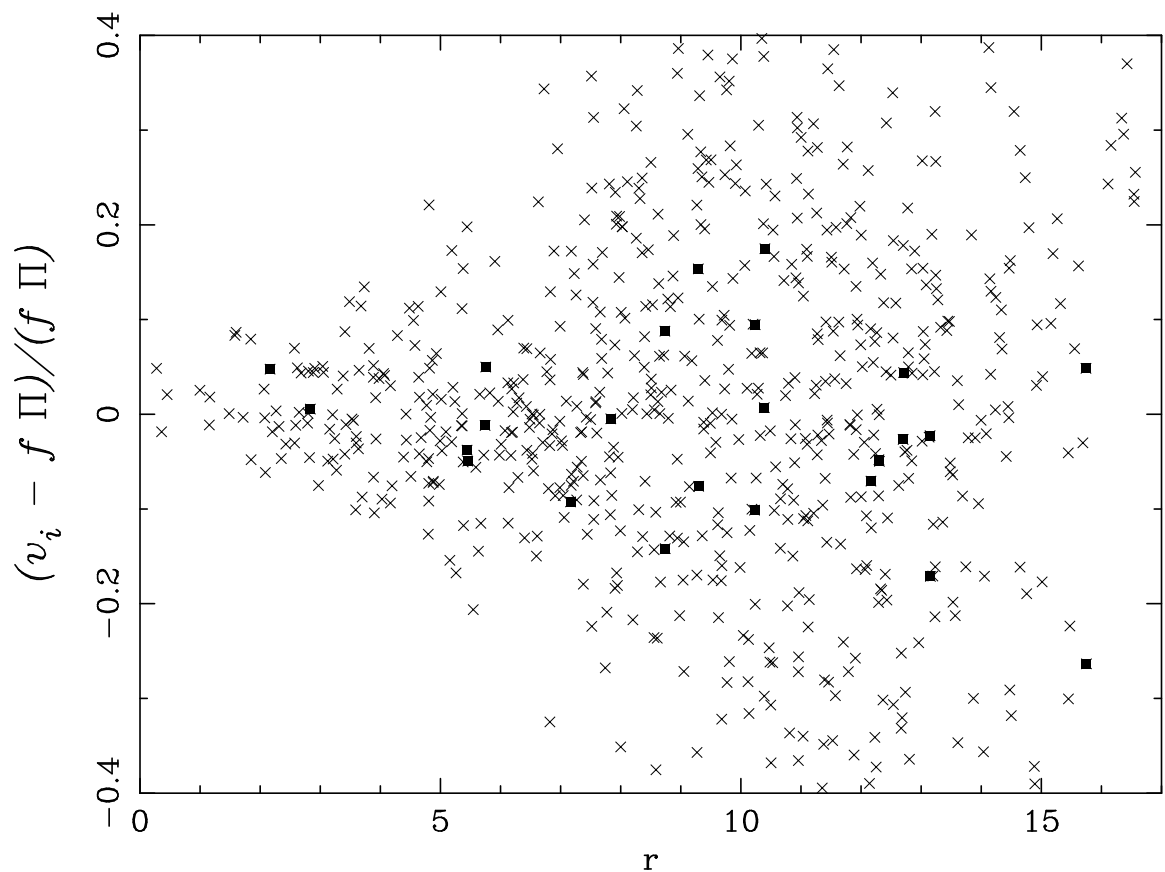}
\caption{a) The residuals resulting from subtracting the fitted ePSF
from the value of each pixel in the science data array plotted
\textit{versus} the distance from the fitted centroid in units of the
ePSF grid spacing.  Only data for objects contributing to the
construction of the ePSF appear.  The slanted crosses are the
residuals for the stars and the solid squares are the residuals for
the QSO for the fiducial image, taken with STIS, of the first epoch
observations of the FOR~J$0238-3443$ field.}
\label{psfresid}
\end{figure}

Another quantitative comparison of the two ePSFs is the $\chi^2$ of the
fit of the ePSF to the science data arrays.
Section~\ref{psfdiag}\ shows that the $\chi^2$ of the QSO does not
differ significantly from those of stars with comparable brightness.
However, the $\chi^2$ is insensitive to systematic trends in the
residuals resulting from subtracting the fitted ePSF from the value of
each pixel in the science data array.  Figure~\ref{psfresid}\ plots
these residuals for the stars (slanted crosses) and QSOs (solid squares and
triangles) contributing to the construction of the ePSF as a function
of distance from the fitted centroid, $r$, in units of the ePSF grid
spacing.  The panels show, from top to bottom, the residuals for the
fiducial images of the first-epoch observations of the fields
FOR~J$0238-3443$ (STIS observations), FOR~J$0240-3434$ (PC observations),
and FOR~J$0240-3438$ (STIS observations).

The distribution of the residuals for the QSO is identical to that for
the stars in the top and middle panels of Figure~\ref{psfresid}.  The
solid squares and triangles in the middle panel correspond to the two
images of the lensed QSO in this field.  In contrast, the distribution
of residuals for the QSO in the bottom panel is different from that for
the stars.  The residuals for the stars are positive near $r = 0$ and
negative near $r=5$, reflecting a small mismatch between the
constructed and true ePSF caused by smoothing.  However, the residuals
for the QSO are negative and the same at both radii.  Therefore, the
image of this QSO is slightly more extended than that of a star.  The
QSO in the FOR~J$0240-3438$ field is the closest of the three, $z =
0.38$ \textit{versus} $z=2.00$ and $1.40$ (Tinney, Da Costa, \&
Zinnecker 1997), and the surrounding galaxy is visible in our images.
The difference between the radial profiles of the QSO in the
FOR~J$0240-3438$ field and that of a star is comparable to the variation
in the ePSF within a field (see Section~\ref{psfdiag}).  For this
reason, the effect on the astrometry is likely to be small.
Ultimately, it will be the agreement between the proper motions derived
from all three fields in Fornax that will determine whether the effect
on the astrometry is indeed negligible.

\section{Determination of the Centroids of Objects}
\label{fitepsf}

	Having determined the ePSF, the algorithm fits for the centroid
of object $j$, implicit in $\vec{\delta}^j$, and its total flux, $f^j$,
by minimizing
\begin{equation}
\chi^{2}(\vec{\delta}^j,f^j)=\sum_{i=1}^{25}
\left(\frac{v_{i}^{j}-f^{j}\Pi(\vec{v}_i-\vec{\delta}^j)}
{u_{i}^{j}}\right)^{2}.
\label{chi2}
\end{equation}
The best-fitting $\vec{\delta}^j$ must be determined numerically,
whereas the best-fitting $f^{j}$ for a given $\vec{\delta}^j$
can be calculated analytically from  
\begin{equation}
f^{j}=\left(\sum_{i=1}^{25}
\frac{v_{i}^{j}\Pi(\vec{v}_i-\vec{\delta}^j)}
{\left(u_{i}^{j}\right)^{2}}\right)/
\sum_{i=1}^{25}\left
(\frac{\Pi(\vec{v}_i-\vec{\delta}^j)}{u_{i}^{j}}\right)^{2}.
\label{flux}
\end{equation}
In both of the above equations, the sum includes only those pixels that
have flags equal to zero.  The algorithm uses bicubic spline
interpolation to evaluate $\Pi(\vec{v}_i-\vec{\delta}^j)$ from the
tabulated ePSF grid values, where the vector $\vec{v}_i-\vec{\delta}^j$
points from the centroid of the object $j$ (which is also the center of
the ePSF grid) to the center of science data pixel $i$.
Equations~\ref{delx} and \ref{dely} determine the centroid of an
object, $\vec{X}\equiv(X_{0},Y_{0})$, from $\vec{\delta}^j$.

The error in a measured centroid has contributions from the noise in
the constructed ePSF, a possible systematic mismatch between the
constructed and true ePSF, and the noise in the science data arrays.
The remainder of this section addresses the importance of these sources
of error by examining: 1) how well the constructed ePSF matches the
science data arrays of objects both as a function of position within
the image and, when averaged over all objects, as a function of
position within the science data array; 2) whether the difference
between the centroids measured for an object in the individual images
of an epoch depend on the the location of the object with respect to
the pixel boundaries in each image; and 3) if the \textit{rms} scatter of the
centroids from the individual images depends on the location of the
object within the field and on the brightness of the object.

The discussion draws on examples taken from the STIS images of the
FOR~J$0238-3443$ field and the PC images of the FOR~J$0240-3434$
field.  However, the conclusions apply to all of our data for Fornax.
The plots presented in this section are for the fiducial first-epoch STIS
or PC images or for all of the first-epoch STIS or PC images.

\subsection{Performance of the Fitting Algorithm: Flux Residual Diagnostics}
\label{psfdiag}

	This section examines the flux residuals resulting from
subtracting the fitted ePSF from the values in the science data array.
A dependence of these flux residuals on location within the field would
indicate a spatially varying ePSF.  If the flux residuals for a star
produce a $\chi^2$ of low probability, then either the uncertainties of
the values in the science data array are underestimated or there is a
statistically significant mismatch between the constructed ePSF and the
true ePSF.  A systematic mismatch could also cause some of the flux
residuals averaged over all objects to be non-zero and to vary within
the science data array.

The left and right panels in Figure~\ref{fluxres}a plot the mean flux
residual divided by its uncertainty,
\begin{equation}
{\cal RF}=\sum_{i=1}^{25}\frac{v_{i}-f\hat{\Pi}(\vec{v}_i-\vec{\delta})}{u_{i}},
\label{fluxreseq}
\end{equation}
for each object in the fiducial STIS image as a function of $X_{0}$ and
$Y_{0}$, respectively.  The solid square is the QSO and the pluses are
the objects contributing to the construction of the ePSF (which include
the QSO for this image).  Figure~\ref{fluxres}b is the corresponding
plot for the PC image.  The solid square and triangle are images A and
B, respectively, of the lensed QSO.  Because ${\cal RF}$ is sensitive
to the sign of the residuals, whereas $\chi^2$ is not, ${\cal RF}$
contains additional information on how the fitted ePSF and the science
data array differ.  For our data, $\cal{RF}$ is a better diagnostic of
variability of the ePSF with position in our fields than is $\chi^2$.

Figure~\ref{fluxres}a shows no discernible correlation between ${\cal
RF}$ and the location of an object in the image.  The mean ${\cal RF}$
is close to zero, though negative, and the departure from zero is
statistically significant.  A difference between the shapes of the
constructed and true ePSF causes the average ${\cal RF}$ to be
non-zero.  For example, Figure~\ref{numbingrid} will show that the
constructed ePSF is broader than the true ePSF in the central region of
the ePSF grid.  The scatter of ${\cal RF}$ around its mean is on the
order of $0.17$ and the size of the scatter does not correlate with
either $X_{0}$ or $Y_{0}$.  The pluses tend to have a larger scatter
than the points because, in Equation~\ref{fluxreseq}, the bright stars
have smaller fractional uncertainties in $v_i$.  They also tend to have
a positive ${\cal RF}$ for $X_0 \lesssim 400$~pixels.  This pattern is
present in all of our STIS data for Fornax to a variable degree and is
particularly conspicuous for the middle epoch of the FOR~J$0240-3438$
field.  We conclude that the ePSF is variable within our STIS images,
though the variability is barely detectable in most of them.

The corresponding plots for the fiducial PC image,
Figure~\ref{fluxres}b, show that ${\cal RF}$ varies with position in
the image.  Unlike Figure~\ref{fluxres}a, the distribution of the
pluses is similar to that of the points.  This similar distribution for
the bright and faint objects is a consequence of the smaller total
fluxes in the PC images: the Poisson noise is larger than the
systematic mismatch between the ePSF and the science data array.  The
typical ${\cal RF}$ for the corners of the image is greater than that
near the center --- again, a pattern exhibited to a variable degree in
all of our PC data for Fornax.  Thus, as for the STIS data, the ePSF
varies within the field --- though the pattern of the variation is
different and the amplitude of the variation is larger.

Figure~\ref{fluxchi}a plots $\chi^{2}$, calculated from
Equation~\ref{chi2}, for each object in the fiducial STIS image as a
function of the flux.  Here the flux is the sum of all values in the
science data array that have flags equal to zero.  Since the summation in
Equation~\ref{chi2} is over 25 science data pixels (less than 25 if
there are pixels with non-zero flags), the $\chi^{2}$ per object should
be around 25.  The figure shows that the value of $\chi^{2}$ is about
25 for those objects with fluxes less than about $3\times 10^{3}$
counts.  For higher fluxes, the $\chi^{2}$ increases with flux.  For
example, the total $\chi^{2}$ for the QSO is several hundred.  Such
large values for $\chi^2$ are not necessarily alarming since, for very
bright objects, the fractional uncertainties in the values in the
science data array are very small and a large $\chi^2$ does not imply
large differences between the array values and the ePSF.

Figure~\ref{fluxchi}b is the corresponding plot of $\chi^2$
\textit{versus} total flux for each object in the fiducial PC image.
The trend of increasing $\chi^2$ with
flux is less apparent because of the lower fluxes and more variable
ePSF in the PC image as compared to the STIS image.

	Figure~\ref{fluxres} shows that the values of $\cal{RF}$ for
the QSOs in fields FOR~J$0238-3443$ and FOR~J$0240-3434$ are similar to
those for the bright stars, which is consistent with the conclusion of
Section~\ref{qpsfspsf} that the images of these QSOs are
indistinguishable from that of a star.  In contrast, the QSO in field
FOR~J$0240-3438$ has a positive value of $\cal{RF}$ that tends to be
larger than that for any bright star.  Its larger value of $\cal{RF}$
is also consistent with the conclusion of Section~\ref{qpsfspsf} that
the radial profile of this QSO is slightly more extended than that of a
star.  For all of the three fields in Fornax, the values of $\chi^{2}$
for the QSO are not markedly different from those for the bright
stars.  Though the $\chi^{2}$ of the QSO is one of the largest, if not
the largest, it is comparable to the $\chi^{2}$ of stars with a similar
flux.  Thus, the $\chi^2$ is not a good indicator of small differences
between the images of a QSO and a star because the $\chi^2$ values of
bright stars are also large.

	Figure~\ref{numbingrid} explores the reason for the large
$\chi^2$ values for bright stars by showing the mean flux residual as a
function of position within the science data array for all of the
STIS images.  The top number in a square tile is the mean over all
objects of the fractional difference between the value of the science
data pixel and the fitted value, defined by
\begin{equation}
\langle \Delta v_{i}/v_{i} \rangle=\frac{1}{N}\sum_{j=1}^{N}
\frac{v_{i}^{j}-f^{j}\Pi(\vec{v}_i-\vec{\delta}^j)}
{f^{j}\Pi(\vec{v}_i-\vec{\delta}^j)}.
\label{delvi}
\end{equation}
Here $j$ is the index of an object and $N$ is the total number of
objects in the sum.  The bottom number in a square tile is the mean
over all objects of the difference between the value of the science
data pixel and the fitted value divided by the uncertainty in the pixel
value, calculated from
\begin{equation}
\langle \Delta v_{i}/u_{i}\rangle = \frac{1}{N}
\sum_{j=1}^{N}\frac{v_{i}^{j}-f^{j}\Pi(\vec{v}_i-\vec{\delta}^j)}
{u_{i}^{j}}.
\label{delviw}
\end{equation}

	The values of $\langle \Delta v_{i}/v_{i} \rangle$ in
Figure~\ref{numbingrid} are positive in the inner $3\times 3$ pixels
and tend to be negative in those outside.  This pattern supports the
idea that the smoothing in the construction of the ePSF causes the
constructed ePSF to be broader than the true ePSF.  In the critical
central $3\times 3$ region, the fractional difference between the ePSF
and true PSF is 5\% or less.  In the less critical outer region, where
the ePSF is small, the difference can be as large as $10\%$.  These
numbers are typical of those for our STIS and PC data for Fornax.

The largest value of $\langle \Delta v_{i}/u_{i} \rangle$ occurs for
the central pixel.  A total of 8538 central pixels contribute to this
average value, so, if there is no difference between the constructed
ePSF and the true ePSF and the uncertainties in the array values are
realistic, then the typical $\langle \Delta v_{i}/u_{i} \rangle$ would
be $\pm$0.011.  Thus, the largest $\langle \Delta v_{i}/u_{i} \rangle$
is statistically significant.  Likewise, the values of $\langle \Delta
v_{i}/u_{i} \rangle$ in the other tiles are statistically significant.
Ideally, the values of $\langle \Delta v_{i}/u_{i} \rangle$ would be
zero, indicating a perfect match between the constructed ePSF and true
ePSF.  The non-zero values again arise from the smoothing imposed during the
construction of the ePSF.  Reducing the smoothing reduces the
systematic differences between the constructed ePSF and true ePSF, but
it increases the random differences.  Both random and systematic
differences lead to errors in the fitted centroids.  We choose the
smoothing which minimizes the total uncertainty in the fitted
centroids.

\subsection{Average Centroids of Objects}

	Let $(X_{0},Y_{0})$ be the coordinates of the centroid of an
object in an image expressed in the coordinate system of the fiducial
image.  If there are $N$ images, there are up to $N$ values of
$(X_{0},Y_{0})$ for a given object -- less if that object is not fitted in
some images.  Let $(\langle X_{0}\rangle, \langle Y_{0}\rangle)$ be the
mean centroid of an object expressed in the coordinate system of the
fiducial image.  The residual of an $(X_{0},Y_{0})$ value for an object
from the $(\langle X_{0}\rangle, \langle Y_{0}\rangle)$ is
\begin{eqnarray}
{\cal RX}&\equiv&\langle X_{0}\rangle-X_{0} \label{rx0} \\
{\cal RY}&\equiv&\langle Y_{0}\rangle-Y_{0} \label{ry0}.
\end{eqnarray}
The \textit{rms} scatter of $X_0$ and $Y_0$ around $\langle X_{0}\rangle$
and $\langle Y_{0}\rangle$, respectively, are measures of how well the
algorithm determines the centroid of the object.

	The positions of objects on the STIS and on the WFPC2 detectors
do not reflect their true positions in the sky due to geometrical
distortion present in the instrument, which makes the image scale vary
across the detector.  The effect of this distortion on the centroid of
an object can be reduced because the relationship between the true
centroid of an object, $(x_{t},y_{t})$, and its measured centroid,
$(x,y)$, with respect to the center of the CCD detector has been
approximately characterized by Brown \etal\ (2002) for STIS and Baggett
\etal\ (2002) for WFPC2 and is given by
\begin{eqnarray}
x_{t}&=&C_{0}+C_{1}x+C_{2}y+C_{3}x^{2}+C_{4}xy+C_{5}y^{2}+C_{6}x^3+C_{7}yx^{2}+
C_{8}xy^{2}+C_{9}y^{3} \\
y_{t}&=&D_{0}+D_{1}x+D_{2}y+D_{3}x^{2}+D_{4}xy+D_{5}y^{2}+D_{6}x^3+D_{7}yx^{2}+
D_{8}xy^{2}+D_{9}y^{3}.
\end{eqnarray}
The second and third columns of Table~2 list the coefficients for the
STIS detector and the fourth and fifth columns do the same for the PC
detector.  These corrections are applied to all of our measured
centroids before calculating the means and residuals described above.
In addition to the above geometrical distortions, the WFPC2 CCD chip has
a ``34-th row defect'' (Shaklan, Sharman, \& Pravdo 1995) which
introduces errors into the measured values of both the $y$ coordinate
of the centroid and the total flux of an object.  Anderson~\&~King
(1999) determined corrections for these errors.  Our algorithm uses
their recipe (their Equation~2) to correct the errors in the $y$
coordinates of the centroids. However, it does not correct the values
of the science data array (their Equation~1) because we think that the
flat-field correction performed by the HST pipeline eliminates the need
for this correction for astrometric purposes.

\subsection{Performance of the Fitting Algorithm: Position Residual
Diagnostics}
\label{positiondiag}

	The two left-hand panels in Figure~\ref{rmssn} plot the
\textit{rms} scatter of $X_0$ (top) and $Y_0$ (bottom) around their
means \textit{versus} the inverse of the $S/N$ for all objects with at
least two measured centroids in the first-epoch STIS images.  The $S/N$
is calculated from Equation~\ref{sigtonoise}.  The two right-hand
panels are the same for objects in the first-epoch PC images.  The
plots in Figure~\ref{rmssn} are typical of our data for Fornax.  The
plots show a nearly linear trend of increasing \textit{rms} scatter
with increasing inverse $S/N$, as is expected from the uncertainties in
the least-squares fit (\textit{e.g.}, Kuijken \& Rich 2002).  The trend
does not go through the origin in any of the plots, demonstrating that
one or more sources of error that are independent of the $S/N$ limit
the accuracy of the fitted centroids for the brightest objects.  The
additional source of error is likely a dependence of the measured
centroid on the location of the true centroid within a pixel -- called
pixel-phase error by Anderson \& King (2000).  Pixel-phase errors arise
because of a mismatch between the constructed and true ePSF.
Intentional degradation of the ePSF by including fewer stars in its
generation produces a larger non-zero intercept in the equivalent of
Figure~\ref{rmssn}.  Even an ePSF constructed from many objects can
have some mismatch if the true ePSF varies across an image and
Figure~\ref{fluxres} shows that this is the case for our Fornax data.
Additional mismatch can arise if the ePSF varies among images at a
given epoch because of a variable size of the guiding jitter.  Our data
has too few stars to model the variations of the ePSF within and
between images without introducing larger errors in the ePSF from
undersampling.  Thus, unavoidable pixel-phase errors cause the minimum
\textit{rms} of about 0.01~pixel, independent of $S/N$, seen in
Figure~\ref{rmssn}.

	The left-hand panels in Figure~\ref{posresid} plot, from top to
bottom, ${\cal RX}$ \textit{versus} the decimal part of the $X$
component of the centroid --- the pixel phase $\Phi_x$, ${\cal RY}$
\textit{versus} the decimal part of the $Y$ component of the centroid
--- the pixel phase $\Phi_y$, ${\cal RX}$ \textit{versus} $\Phi_y$, and
${\cal RY}$ \textit{versus} $\Phi_x$.  All of the panels are for
objects contributing to the construction of the ePSF for the
first-epoch STIS images.  The right-hand panels show the same
quantities for objects in the first-epoch PC images.  The plots of
${\cal RX}$ \textit{versus} $\Phi_x$ and ${\cal RY}$ \textit{versus}
$\Phi_y$ for the STIS images probably show systematic trends in the
residuals that reflect pixel-phase errors with an amplitude of about
0.01~pixel.  The points for the QSO show the pixel-phase errors most
clearly: ${\cal RX}$ tends to be negative for $\Phi_x \lesssim 0.5$ and
positive above 0.5 and the opposite pattern is present for ${\cal RY}$
\textit{versus} $\Phi_y$.  The two plots of the cross-dependences do
not show any trends.

The plots for the PC images show no clear evidence for pixel-phase
errors.  However, if these errors depend on the position within an
image, as they might because of the variation of the ePSF for the PC
images implied by Figure~\ref{fluxres}, a plot for the entire region of
all images will tend to show scatter instead of a trend.

\subsection{Uncertainty of the Average Centroid}

The uncertainty of the average centroid is the \textit{rms} scatter
around the mean divided by the square root of the effective number of
measurements.  Estimating the uncertainty of the average centroid is
complicated by the presence of pixel-phase errors.  When these errors
make the largest contribution to the uncertainty of the measured
centroid, as they do for the brightest few objects in our sample (see
Figure~\ref{rmssn}), the effective number of measurements is the number
of dither positions rather than the total number of measurements of the
centroid.  Because pixel-phase errors cannot be eliminated from our
data, we have chosen to include their effect on the uncertainty of the
mean centroid, $\sigma_{\langle X_0\rangle}$ or
$\sigma_{\langle Y_0\rangle}$, with the ad-hoc formula
\begin{equation}
\sigma_{\langle X_0\rangle,\langle Y_0\rangle} = \frac{\sigma_{X_0,Y_0}}
{\left(\left(\sigma_0/\sigma_{X_0,Y_0}\right)^2 N_{d} +
\left(1 - \left(\sigma_0/\sigma_{X_0,Y_0}\right)^2\right) N_{obs}
\right)^{1/2}}.
\label{sigmaavg}
\end{equation}
Here $\sigma_{X_0,Y_0}$ is the \textit{rms} scatter around
$\langle X_0\rangle$ or $\langle Y_0\rangle$, $\sigma_0$ is the minimum
uncertainty due to pixel-phase errors, $N_{d}$ is the number of dither
positions, and $N_{obs}$ is the number of measurements of the
centroid.  From Figure~\ref{rmssn}, the value of $\sigma_0$ is 0.013.
The value of $\sigma_0$ ranges from 0.012 to 0.013 for the Fornax data.

\section{Determination of Proper Motion and Space Velocity}
\label{measurepm}

\subsection{The Measured Proper Motion}
\label{measurepm2}

	The average centroid of an object at a given epoch is in the
	coordinate system of the fiducial image for this epoch.  The
centroid differs between epochs both because of changes in the pointing
of the telescope and the proper motion of the dSph.  Determining the
proper motion from centroids of objects at different epochs requires
adopting a standard coordinate system tied to the stars of the dSph.
These stars determine a transformation between the coordinate system of
the fiducial image at the later epoch and the fiducial image at the
earlier epoch.

	The transformation to the standard coordinate system has
a similar form to the one given in Equation~\ref{tran}:
\begin{equation}
\vec{X}^{tr}(t_2)=s R(\theta)\times
\left(\vec{X}(t_2) -\vec{X_{Q}}(t_2)\right) - \vec{X_{Q}}^{tr}(t_2).
\label{tran2}
\end{equation}
The transformation coefficients result from varying the plate scale,
$s$, angle, $\theta$, and the transformed coordinates of the QSO,
$(X_{Q}^{tr}(t_2),Y_{Q}^{tr}(t_2))$, to minimize
\begin{equation}
\chi^2=\sum_{i=1}^{N}\left(\frac{\left(X_{i}(t_{1})-
X_{i}^{tr}(t_{2})\right)^{2}}{\sigma^{2}_{X_{i}(t_{1})}+
\sigma^{2}_{X^{tr}_{i}(t_{2})}} +
\frac{\left(Y_{i}(t_{1})-
Y_{i}^{tr}(t_{2})\right)^{2}}{\sigma^{2}_{Y_{i}(t_{1})}+
\sigma^{2}_{Y^{tr}_{i}(t_{2})}}
\right).
\label{chi2tr}
\end{equation}
Here the uncertainties of the transformed centroids are given by
\begin{eqnarray}
\sigma_{X^{tr}_{i}(t_{2})} &=& s\left(\cos^{2}(\theta)\,
\sigma^{2}_{X_{i}(t_{2})}+
\sin^{2}(\theta)\,\sigma^{2}_{Y_{i}(t_{2})}   \right)^{1/2}
\label{sigxtr} \\
\sigma_{Y^{tr}_{i}(t_{2})} &=& s\left(\sin^{2}(\theta)\,
\sigma^{2}_{X_{i}(t_{2})}+
\cos^{2}(\theta)\,\sigma^{2}_{Y_{i}(t_{2})}   \right)^{1/2}.
\label{sigytr}
\end{eqnarray}
The objects that determine the transformation are those with a $S/N$
above some limit, which is 20 for the Fornax data.  The QSO is excluded
from the determination of the transformation, as is every object whose
change in position between epochs is more than 2.5 times, for STIS, or
4.5 times, for the PC, its uncertainty.  The set of excluded objects is
determined iteratively, with the iteration ceasing when the set stops
changing.  The number of excluded objects is at most 15\% of a sample
of 60 or more objects.  The final proper motion does not depend
sensitively on the limits that determine the sample which defines the
transformation.  A typical change in scale between epochs is a few
parts per 100,000 and the typical angle of rotation is on the order of
a few thousandths of a degree or less.

The uncertainty in the transformed coordinates,
$(X^{tr}(t_2),Y^{tr}(t_2))$,  depends on both the uncertainty in the
measured centroid, $(X(t_2),Y(t_2))$, and on the uncertainty in the
transformation.  For the QSO, these uncertainties are independent and
should be added in quadrature.  Even for the objects that contributed to
the determination of the transformation, adding the two uncertainties
in quadrature is a reasonable procedure as long as the sample of objects
is at least a few tens.

The contribution to the uncertainty in $(X^{tr}(t_2),Y^{tr}(t_2))$ from
the uncertainty in the coefficients of the transformation could be
estimated by varying the coefficients so that $\chi^2$ increases from its
minimum value by one --- see the discussion in Chapter~15 of
Press \etal\ (1992).  Because implementing this estimate is very complex,
we have adopted a simpler and more intuitive approach.  The values of
$(X_{Q}^{tr}(t_2), Y_{Q}^{tr}(t_2))$ that minimize $\chi^2$ are
\begin{eqnarray}
X_{Q}^{tr}(t_2) &=& \left(\sum_{i=1}^{N}
\frac{X_i(t_1) - s\left((X_i(t_2)-X_Q(t_2))\cos(\theta) -
(Y_i(t_2)-Y_Q(t_2))\sin(\theta)\right)}
{\sigma_{X_i(t_1)}^2 + \sigma_{X_i^{tr}(t_2)}^2} \right) / \nonumber \\
 & & \hspace{1.0truein} \left(\sum_{i=1}^{N} \frac{1}
{\sigma_{X_i(t_1)}^2 + \sigma_{X_i^{tr}(t_2)}^2} \right)
\label{xqtr}
\end{eqnarray}
and
\begin{eqnarray}
Y_{Q}^{tr}(t_2) &=& \left(\sum_{i=1}^{N}
\frac{Y_i(t_1) - s\left((X_i(t_2)-X_Q(t_2))\sin(\theta) -
(Y_i(t_2)-Y_Q(t_2))\cos(\theta)\right)}
{\sigma_{Y_i(t_1)}^2 + \sigma_{Y_i^{tr}(t_2)}^2} \right) / \nonumber \\
 & & \hspace{1.0truein} \left(\sum_{i=1}^{N} \frac{1}
{\sigma_{Y_i(t_1)}^2 + \sigma_{Y_i^{tr}(t_2)}^2} \right).
\label{yqtr}
\end{eqnarray}
Applying propagation of errors to the above equations yields the
following uncertainties for the two parameters:
\begin{eqnarray}
\sigma_{X_{Q}^{tr}(t_2)} &=& \left(\sum_{i=1}^{N}\frac{1}
{\sigma_{X_i(t_1)}^2 + \sigma_{X_i^{tr}(t_2)}^2} \right)^{-1/2}
\label{sigxqtr}\\
\sigma_{Y_{Q}^{tr}(t_2)} &=& \left(\sum_{i=1}^{N} \frac{1}
{\sigma_{Y_i(t_1)}^2 + \sigma_{Y_i^{tr}(t_2)}^2} \right)^{-1/2}.
\label{sigyqtr}
\end{eqnarray}
The above estimates for the contribution from the transformation to the
uncertainty in $(X^{tr}(t_2), Y^{tr}(t_2))$ are accurate if the
covariances of the parameters in the transformation are small, which is
the case if the number of objects determining the transformation is
much larger than the number of parameters and the objects are widely
distributed in the field.  The data for Fornax satisfies these
requirements.  The uncertainties in the transformed coordinates of
objects other than the QSO also have contributions from the
uncertainties in $s$ and $\theta$.  We neglect these additional
contributions in the calculation of the uncertainties since they do not
directly affect the measured proper motion of the dSph.  Thus, the
uncertainty in the centroid from the later epoch transformed into the
coordinate system of the first epoch is the measurement uncertainty of
the later-epoch centroid added in quadrature to the uncertainty arising
from the transformation, given by Equations~\ref{sigxqtr} and
\ref{sigyqtr}.

As defined in Section~\ref{overview}, $p_x$ and $p_y$ are the
components of the difference between the two centroids of an individual
object, measured at different epochs, in the standard coordinate
system.  The two panels in Figure~\ref{sigvssn} plot the uncertainty in
$p_{x}$, $\sigma_{p_x}$, (top) and the uncertainty in $p_{y}$,
$\sigma_{p_y}$, (bottom) \textit{versus} the inverse of the average
$S/N$.  The figures in this section use the epoch 2000 and 2001 data
for the FOR~J$0238-3443$ field.  The $S/N$ is from the earlier epoch.
The plots show the expected approximate linear increase of
$\sigma_{p_{x}}$ and $\sigma_{p_{y}}$ with decreasing $S/N$.  The
smallest uncertainty in the change in a centroid ($p_x$ or $p_y$) is
about 0.005~pixel.  This uncertainty is comparable to that necessary to
measure the proper motion of a typical dSph, as estimated in
Section~\ref{intro}.

	The panels in Figure~\ref{pxpyxy} show how the values of
$p_{x}$ and $p_{y}$ depend on location for objects with a $S/N$ greater
than 30.  Figure~\ref{sigvssn} shows that this limit corresponds to an
uncertainty in $p_{x}$ or $p_{y}$ of less than approximately
0.01~pixel.  The two left-hand panels plot $p_{x}$ (top) and $p_{y}$
(bottom) \textit{versus} $X(t_{1})$.  The two right-hand panels
similarly plot the same quantities {\it versus} $Y(t_{1})$.  None of
the panels in Figure~\ref{pxpyxy} shows a trend with either $X(t_{1})$
or $Y(t_{1})$.  The absence of trends argues that  the form of the
transformation to the standard system is adequate and that any
variation of the ePSF across the image does not affect significantly
$p_{x}$ and $p_{y}$ and, thus, does not affect the measured proper
motion.

A figure analogous to Figure~\ref{pxpyxy} for the PC data for field
FOR~J$0240-3434$ shows that $p_x$ and $p_y$ depend on $X(t_{1})$ and
$Y(t_{1})$, respectively, with a pattern that resembles the one for
$\cal{RF}$ in Figure~\ref{fluxres}b.  The dependence of $p_x$ and $p_y$
on location argues that the ePSF changed enough between epochs to
introduce systematic errors in the proper motion.  Adding terms
depending on $X^2$ and $Y^2$ to the transformation,
Equation~\ref{tran2}, reduces, but does not completely eliminate this
dependence.  The two values for the $X$~component of the proper motion
derived using the two images of the QSO in this field differ by about
twice the uncertainty in their difference.  The two $Y$~components
agree within their uncertainties.  We increase the uncertainties in
both the $X$ and $Y$ components of the measured proper motions for this
field to reflect the remaining systematic errors.  The uncertainties
are equal to each other and have a value that makes the difference in
the two measured $X$~components of the proper motion equal to its
uncertainty.

	Figure~\ref{pxvspy} plots $p_{y}$ {\it versus} $p_{x}$ for the
same objects as in Figure~\ref{pxpyxy}.  The distribution of points is
centered on and isotropic about the origin, as is expected for stars of
Fornax.  The solid square representing the QSO is clearly offset from
the origin and its error bars indicate that this offset is about twice
its uncertainty.  Figure~\ref{pxvspy2}, which plots
$p_{y}/\sigma_{p_{y}}$ \textit{versus} $p_{x}/\sigma_{p_{x}}$, shows
that the QSO has one of the most significant offsets from the origin.
A $2\sigma$ offset is not a significant detection of the proper motion
of Fornax, however these data have a time baseline of only one year.
Our measurements for three fields, two of which have baselines of two
years, do produce a significant detection.

	Multiplying $-p_{\textrm{\tiny Q},x}$ and $-p_{\textrm{\tiny
Q},y}$ by the image scale and dividing by the time baseline between the
two epochs yields the components of the proper motion of Fornax in the
standard coordinate system, $\mu_{x}$ and $\mu_{y}$.  Rotating the
coordinate system using the known position angle of the $Y$ axis of the
standard coordinate system gives the measured
$\mu_{\alpha}$ and $\mu_{\delta}$
in the equatorial coordinate system.  Table~3 gives these measured
quantities for the three fields in the direction of the Fornax dSph in
units of milliarcseconds per century.  These components of the proper
motion are those that are measured on the sky without applying
corrections for the motions of the Sun and the LSR.  The following
section applies the appropriate corrections and uses the results to
calculate the velocity of Fornax in various reference frames.

\subsection{Galactic-Rest-Frame Proper Motion in the Equatorial Coordinate
System}

	The measured proper motion includes the effects of the motion
of the LSR and the peculiar motion of the Sun with respect to the LSR.
Both motions must be removed from the measured proper motion to obtain
the proper motion of a dSph in the Galactic rest frame.  Let
$(\mu_{\alpha}^{\mbox{\tiny{LSR}}}, \mu_{\delta}^{\mbox{\tiny{LSR}}})$
and $(\mu_{\alpha}^{\odot}, \mu_{\delta}^{\odot})$ be the contributions
to the measured proper motion from the motion of the LSR and the
peculiar motion of the Sun, respectively.  The components of the
rest-frame proper motion of the dSph in the equatorial coordinate
system are then given by
\begin{eqnarray}
\mu_{\alpha}^{\mbox{\tiny{Grf}}}&=&\mu_{\alpha}-
\mu_{\alpha}^{\mbox{\tiny{LSR}}}-\mu_{\alpha}^{\odot} \\
\mu_{\delta}^{\mbox{\tiny{Grf}}}&=&\mu_{\delta}-
\mu_{\delta}^{\mbox{\tiny{LSR}}}-\mu_{\delta}^{\odot}.
\end{eqnarray}
The corrections due to the motion of the LSR are (Mihalas \& Binney 1981)
\begin{eqnarray}
\mu_{\alpha}^{\mbox{\tiny{LSR}}}&=&\phantom{-}\frac{\Theta_0
\sin (\lambda) \sin(\psi)}{4.741\, d} \label{mucor1} \\
\mu_{\delta}^{\mbox{\tiny{LSR}}}&=&-\frac{\Theta_0 \sin(\lambda) \cos(\psi)}
{4.741\, d}.
\label{mucor2}
\end{eqnarray}
Here $\Theta_0$ is the circular velocity of the
LSR with respect to the Galactic center, 220~km~s$^{-1}$, $\lambda$ is the
angular distance between the dSph and the apex of the motion of the LSR, and
$\psi$ is the angle between two great circles, one passing through the
dSph and the North Celestial Pole (NCP) and the other passing through
the dSph and the apex.  The quantity $d$ is the heliocentric distance
to the dSph in pc.

	Replacing $\Theta_0$ with
$\mathcal{V}_{\odot}=(u_{\odot}^{2}+v_{\odot}^{2}+w_{\odot}^{2})^{\frac{1}{2}}$,
where $(u_{\odot}, v_{\odot}, w_{\odot})$ are the components of the
solar motion with respect to LSR, and redefining the angles $\lambda$
and $\psi$ to be with respect to the apex of the solar motion,
equations \ref{mucor1} and \ref{mucor2} also give
$(\mu_{\alpha}^{\odot}, \mu_{\delta}^{\odot})$.  The $u_{\odot}$
component of the solar motion is positive if it points radially away
from the Galactic center; the $v_{\odot}$ component is positive if it
is in the direction of rotation of the Galactic disk; and the
$w_{\odot}$ component is positive if it is in the direction of the
North Galactic Pole (NGP).  In this work $(u_{\odot},v_{\odot},
w_{\odot})=(-10.0,5.25,7.17) \pm (0.36,0.62,0.38)$~km~s$^{-1}$
(Dehnen~\&~Binney 1998).

\subsection{Galactic Coordinate System}

	The components of the proper motion in the galactic coordinate
system, $(\mu_{l}, \mu_{b})$, are
\begin{eqnarray}
\mu_{l}&=&\phantom{-}\mu_{\alpha}\cos (\eta) + \mu_{\delta} \sin (\eta) \\
\mu_{b}&=&-\mu_{\alpha}\sin (\eta) + \mu_{\delta} \cos (\eta),
\label{mulmub}
\end{eqnarray}
where $\eta$ is an angle between two great circles intersecting at the
location of dSph on the sky.  One great circle passes through the dSph
and the NCP and the other great circle passes
through the dSph and the NGP.  The value of $\eta$ can be calculated
from a spherical triangle whose vertices are at the dSph, NCP, and NGP
using standard spherical trigonometry. The sign of $\eta$ is
positive if, upon going around this triangle from the dSph through
the NCP and the NGP, the triangle
is on the left (Smart 1977).

\subsection{Space Velocity of a dSph in the LSR and Galactic Rest
Frames}

	Let $(u, v, w)$ be the components of the space velocity of a
dSph with respect to the LSR.  In this coordinate system, $u$ is
positive if it points radially away from the Galactic center; $v$ is
positive if it is in the direction of Galactic rotation; and $w$
is positive if it is in the direction of the NGP.  These directions
are all defined at the location of the LSR.  The $(u, v, w)$
components are given by
\begin{eqnarray}
u&=&-v_{r} \cos (b) \cos (l) + \mu_{b} d \sin
(b) \cos (l) + \mu_{l} d \sin (l) + u_{\odot} \\ 
v&=&\phantom{-}v_{r} \cos (b) \sin (l) - \mu_{b} d \sin (b) \sin (l) +
\mu_{l} d \cos (l) + v_{\odot} \\ 
w&=&\phantom{-}v_{r} \sin (b) + \mu_{b} d \cos (b)
+ w_{\odot}.
\end{eqnarray}
Here $v_{r}$ is the heliocentric radial velocity of the dSph.

	We define a similar cylindrical coordinate system at the
location of the dSph to express the components of the velocity of the
dSph with respect to the Galactic center, $(\Pi, \Theta, Z)$.  In this
coordinate system, $\Pi$ is parallel to the plane of the Galactic disk
and is positive in the direction away from the Galactic center.  The
velocity components are given by,
\begin{eqnarray}
\Pi&=&\phantom{-}u \cos (\beta) + (v + \Theta_0) \sin (\beta) 
\label{vpi}\\
\Theta&=&-u \sin (\beta) + (v + \Theta_0) \cos (\beta) 
\label{vtheta}\\
Z&=& \phantom{-}w, \label{vz}
\end{eqnarray}
where $\beta$ is the angle between the line connecting the Galactic
center and the origin of the LSR and the line connecting the Sun
and a point corresponding the projection of the dSph onto the
Galactic plane.

\section{Results}

	Table~3 lists the measured proper motion of the Fornax dSph
galaxy determined in each of three fields.  The two measurements for the
FOR~J$0240-3434$ field are for the two images of the lensed QSO.  The
three measurements for the three pairs of epochs for the
FOR~J$0240-3438$ field have the same set of high proper motion stars
rejected in the determination of the coordinate transformation between
epochs.  The first and last epoch determine the set of stars rejected
since this pair of epochs produces the most accurate proper motions.
The first column of the table gives the name of the field; the second
column gives the two epochs for which the third and fourth columns give
the components in the J2000.0 equatorial coordinate system of the
measured proper motion in mas~cent$^{-1}$:  $\mu_{\alpha}$ and
$\mu_{\delta}$, respectively.  These values of the proper motion do not
include any corrections for the solar motion or for the motion of the
LSR; they are as-measured on the sky.  The bottom line of Table~3 gives
the weighted average of $\mu_{\alpha}$ and of $\mu_{\delta}$ and their
corresponding uncertainties.

	The values of $\mu_{\alpha}$ and $\mu_{\delta}$ for the
FOR~J$0238-3443$, FOR~J$0240-3434$, and FOR~J$0240-3438$ fields
represent independent measurements of the proper motion of Fornax.  The
two measured proper motions for the two images of the lensed QSO in the
FOR~J$0240-3434$ field are not completely independent since the
measurements share the transformation into the fiducial coordinate
system.  However, the uncertainties in the $p_x$ and $p_y$ of each
image of the lensed QSO are much larger than the uncertainties in the
transformation, so we consider these two measurements to be
independent.

The tabulated uncertainties for $\mu_{\alpha}$ and $\mu_{\delta}$ for
the FOR~J$0240-3434$ field are about twice as large as those given by
the uncertainty in $p_{\textrm{\tiny Q},x}$ and $p_{\textrm{\tiny
Q},y}$ to account for the remaining dependence of $p_x$ and $p_y$ on
location.  The tabulated uncertainties result from the uncertainties in
$\mu_x$ and $\mu_y$ described in Section~\ref{measurepm2}.

\subsection{Galactic-Rest-Frame Proper Motion of Fornax}

	The third and fourth columns of Table~4 give the proper motion
of Fornax in the Galactic rest frame, $\mu_{\alpha}^{\mbox{\tiny{Grf}}}$
and $\mu_{\delta}^{\mbox{\tiny{Grf}}}$ respectively, determined in the
three fields.  This rest-frame proper motion assumes a distance to
Fornax of 138~kpc (Mateo 1998).  The first two columns of Table~4 are
the same as those for Table~3.  The corresponding values of
$\mu_{l}^{\mbox{\tiny{Grf}}}$ and $\mu_{b}^{\mbox{\tiny{Grf}}}$ are in
the fifth and and sixth columns, respectively.  The next three columns
are the $\Pi$, $\Theta$, and $Z$ components of the velocity of Fornax
with respect to the Galactic center (see Equations~\ref{vpi},
\ref{vtheta}, and \ref{vz}).  These values additionally assume a
distance between the Sun and the Galactic Center of 8.5~kpc and a
heliocentric radial velocity for Fornax of $53 \pm 3$~km~s$^{-1}$
(Mateo 1998).  The final two columns give the radial and tangential
velocities of Fornax with respect to the Galactic center.  The bottom
line of Table~4 gives the weighted average for the tabulated quantities
and their corresponding uncertainties.

The three values for the rest-frame proper motion of Fornax derived
from the three pairs of epochs available for the FOR~J$0240-3438$ field
agree within their uncertainties.  The remainder of the article uses
the value of the proper motion for the longest baseline.  We will
derive the proper motion using all three epochs for each field once we
have the complete set of data for Fornax.
 
The FOR~J$0238-3443$ and FOR~J$0240-3434$ fields and the
2000$\rightarrow$2002 baseline for the FOR~J$0240-3438$ field yield
four independent measurements of the proper motion of Fornax.  The
weighted mean of the magnitudes of the four rest-frame proper motion
vectors is $48 \pm 13$~mas~cent$^{-1}$.  The chi-square of the scatter
around this mean is 0.41 for 3 degrees of freedom, a value which should
be exceeded 94\% of the time by chance.  The weighted mean position
angle of the rest-frame proper motion vector and its uncertainty are
$145 \pm 15$~degrees.  The $\chi^2$ of the scatter around this mean is
1.3 for 3 degrees of freedom, a value which should be exceeded 73\% of
the time by chance.  Thus, the four independent measurements of the
proper motion agree within their uncertainties.

The position angle of the major axis of Fornax is $48\pm 6$~degrees
(Mateo 1998).  Thus, the rest-frame proper motion is nearly
along the minor axis.  The average Galactocentric radial and tangential
velocities in Table~4 show that the orbital motion of Fornax is nearly
across our line of sight.  Thus, Fornax is not elongated along
its orbit.

\subsection{Galactic Orbit of Fornax}
\label{orbit}

The values for the average $\Pi$, $\Theta$, and $Z$ in Table~4 show
that Fornax has a retrograde and moderately inclined Galactic orbit.
The small size of the Galactocentric radial velocity compared to the
tangential velocity implies that Fornax is near apogalacticon or
perigalacticon.  We estimate a value of the apo- and perigalacticon for
the logarithmic potential of the Milky Way given by
\begin{equation}
\Phi(R)= 
\frac{1}{2}\Theta_{0}^{2}\ln \left ( \frac{R}{R_{0}}\right ),
\label{gpot}
\end{equation}
where $R_{0}$ is the current Galactocentric radius of Fornax.  Using
the conservation of energy and of angular momentum, the above equation
can be rewritten as
\begin{equation}
\ln \left (\frac{R_{m}}{R_{0}}\right) =\left
(\frac{V_{r}}{\Theta_{0}}\right)^{2} +
\left(\frac{V_{t}}{\Theta_{0}}\right)^{2}\left(1-\left
(\frac{R_{0}}{R_{m}}\right)^{2}\right).
\label{apoperi}
\end{equation}
Here $V_{r}$ and $V_{t}$ are the current Galactocentric radial and
tangential velocities of Fornax, respectively, and $R_{m}$ is either
the apogalacticon, $R_{a}$, or the perigalacticon, $R_{p}$.  For
$V_{r}=-40$~km~s$^{-1}$ and $V_{t}=310$~km~s$^{-1}$, $R_{p}=0.99
R_{0}$ and $R_{a}=7.2 R_{0}$.  Such a large apogalacticon argues that
Fornax is not bound to the Milky Way.  Note, however, that decreasing
$V_{t}$ by $1\sigma$ reduces $R_a$ to $2.6 R_{0}$, while $R_p$ remains
practically unchanged, and that a circular orbit ($V_{r} = 0$,
$V_{t} = 220$~km~s$^{-1}$) is consistent with the measured values.
The plausible range of values for $V_{r}$ and $V_{t}$
imply a space velocity larger than the assumed Galactic circular
velocity, demonstrating that Fornax is close to perigalacticon.

\subsection{A Lower Limit for the Mass of the Milky Way}

	If Fornax is bound gravitationally to the Milky Way, its
Galactocentric velocity provides a lower limit on the mass of the Galaxy.
If the gravitational potential
of the Milky Way is spherically symmetric, then the lower limit on its
mass, $\mathcal{M}$, is
\begin{equation}
\mathcal{M}=\frac{R\,(V_{r}^{2}+V_{t}^{2})}{2G},
\label{minmass}  
\end{equation}
where $R$ is the Galactocentric radius of the dSph.  Substituting in
the average values for the velocities and $R = 140$~kpc gives
$\mathcal{M} = (1.6 \pm 0.8) \times 10^{12}~\mathcal{M}_{\odot}$.  The
stated uncertainty in this limit comes from the uncertainty in the
velocity of Fornax.

\subsection{Is Fornax a Member of a Stream?}

The location of the Draco, Sculptor, and Ursa Minor dSphs on the great
circle of the Magellanic Stream (Lynden-Bell 1976; Kunkel \& Demers
1977), the alignment of the major axes of Draco and Ursa Minor with
this great circle (Lynden-Bell 1983), and the possible correlation of
the positions of other dSphs and globular clusters with and the
alignment of their major axes along great circles in the sky
(\textit{e.g.}, Lynden-Bell 1994, Majewski 1994, Lynden-Bell \&
Lynden-Bell 1995) suggest the presence of coherent streams in the
Galactic halo.  Lynden-Bell \& Lynden-Bell (1995) proposed membership
for Fornax in four possible streams and predicted its proper motion for
each case.  Since the other proposed members of these streams are at
smaller Galactocentric radii than Fornax, the large perigalacticon of
Fornax found in Section~\ref{orbit} argues against the reality of these
four streams.  The predicted proper motions are in the same quadrant as the
average value in Table~3, however, the magnitude of the measured proper
motion vector is larger than the predictions by factors of 1.7 to 3.8.
The discrepancy between the predicted and our measured proper motion is
more than 2$\sigma$.

Lynden-Bell (1982; see also Majewski 1994) also proposed that Fornax
could be a member of a stream with the Leo~I, Leo~II, and Sculptor
dSphs.  The direction of our measured rest-frame proper motion is
52~degrees from the great circle containing Fornax, Leo~I, and Leo~II.
The discrepancy between the directions is more than three times the
uncertainty in the position angle of the rest-frame proper motion
vector.

\section{Summary}

	This article describes our method for measuring proper motions
of dSph galaxies from dithered images taken with HST at multiple
epochs, each of which contains at least one previously-known QSO.  The
steps in this method are: construct an ePSF from the bright objects in
the images, fit the ePSF to all of the objects to determine their
accurate centroids, and transform the centroids to a common coordinate
system.  The measured proper motion of a dSph is the average shift of
the positions of its stars with respect to the QSO.

	The construction of a well-sampled ePSF is central to our
method.  Ideally, the ePSF would have the freedom to vary from image to
image and to vary within an image.  However, because there are
relatively few stars with sufficiently high $S/N$ in our fields,
constructing a well-sampled ePSF requires using a single, constant ePSF
for a field at each epoch.  This limitation introduces an error that
depends on the location of the centroid within a pixel and limits the
accuracy of a measured centroid to 0.013~pixel for a single image.
Averaging the centroids from images at eight independent dither
positions yields a proper motion accurate to about 35~mas~cent$^{-1}$
for a pair of epochs separated by 1~year.  This accuracy is sufficient
to measure the proper motion of the Fornax dSph galaxy.

	We present four independent measurements of the proper motion
of Fornax based on data from three different fields taken with two
different detectors --- STIS and PC.  These measurements agree within
their uncertainties, lending credibility to the method and the measured
value.

The average measured proper motion of Fornax is $\mu_{\alpha} = 49 \pm
13$~mas~cent$^{-1}$ and $\mu_{\delta} = -59 \pm 13$~mas~cent$^{-1}$.
The average proper motion vector corrected for the motion of the Sun
and LSR (the proper motion in the Galactic rest frame) has a magnitude
of $48 \pm 13$~mas~cent$^{-1}$ and a position angle of $145 \pm
15$~degrees.  This position angle places the rest-frame proper motion
vector approximately along the minor axis of Fornax.  Fornax has
Galactocentric radial and tangential velocities of $-40 \pm
50$~km~s$^{-1}$ and $310 \pm 80$~km~s$^{-1}$, respectively.  Assuming
that the Galaxy has a flat rotation curve with a circular velocity of
220~km~s$^{-1}$, Fornax is near perigalacticon and, unless its space
velocity is about 1$\sigma$ or more below the measured value, it
reaches such large Galactocentric radii that it is bound to the Local
Group rather than to the Galaxy.  If Fornax is bound to the Galaxy, the
implied lower limit on the mass of the Milky Way is $(1.6 \pm 0.8)
\times 10^{12}~\mathcal{M}_{\odot}$.

Our measured proper motion for Fornax precludes its membership in the
proposed Fornax-Leo-Sculptor stream or in any of the streams proposed in
Lynden-Bell \& Lynden-Bell (1995).

\acknowledgments

CP and SP acknowledge the financial support of the Space Telescope
Science Institute through the grants HST-GO-07341.03-A and
HST-GO-08286.03-A.  EWO acknowledges support from the Space Telescope
Science Institute through the grants HST-GO-07341.01-A and
HST-GO-08286.01-A and from the National Science Foundation through the
grants AST-9619524 and AST-0098518.  MM acknowledges support from the
Space Telescope Science Institute through the grants HST-GO-07341.02-A
and HST-GO-08286.02-A and from the National Science Foundation through
the grant AST-0098661.  HM acknowledges support from the National
Science Foundation through the grant AST-0098435.  DM is supported by
FONDAP Center for Astrophysics 15010003.

\clearpage

\newpage

\newpage

\newpage

\newpage

\newpage

\newpage

\clearpage
\setcounter{figure}{6}
\begin{figure}[p]
\centering
\includegraphics[angle=-90,scale=0.85]{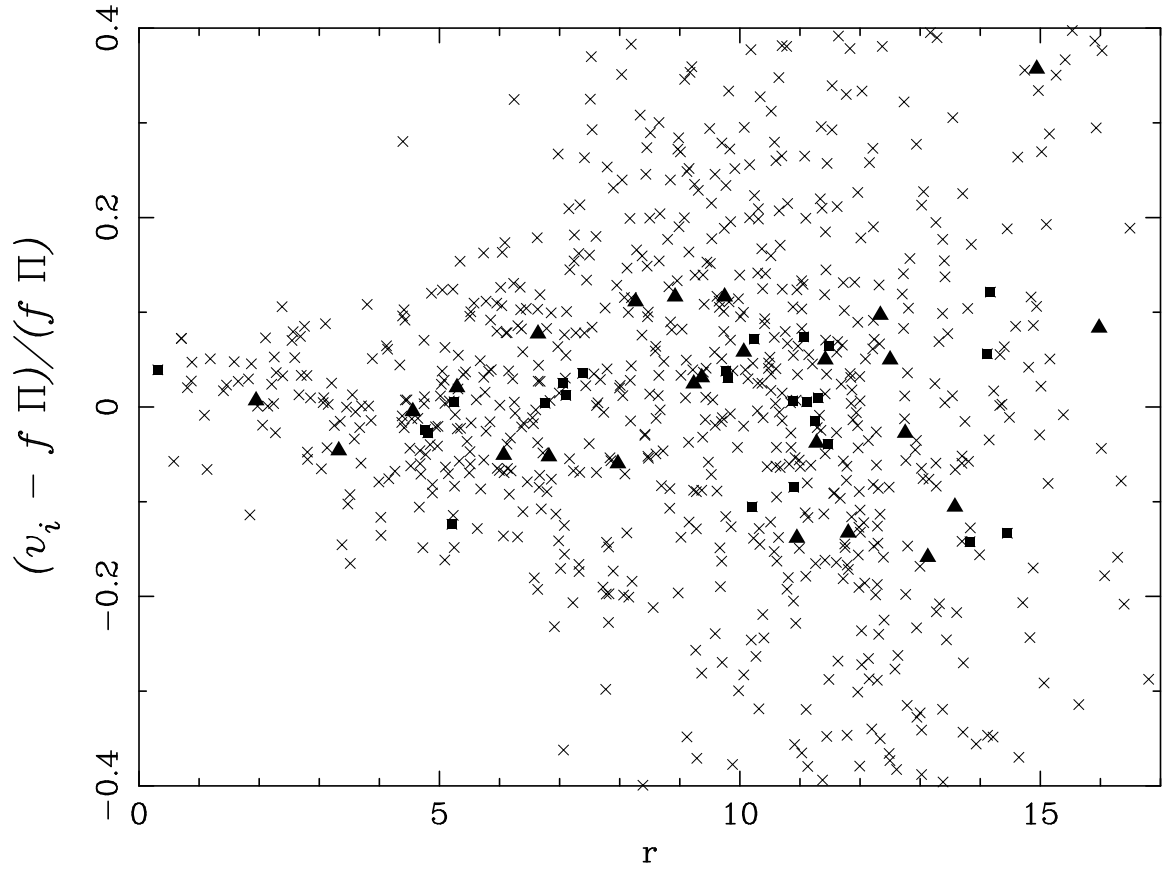}
\caption{b) The same as a) for the
fiducial image, taken with the PC, for the FOR~J$0240-3434$ field.
The solid triangles represents the points for the second image of the
lensed QSO.}
\end{figure}

\clearpage
\setcounter{figure}{6}
\begin{figure}[p]
\centering
\includegraphics[angle=-90,scale=0.85]{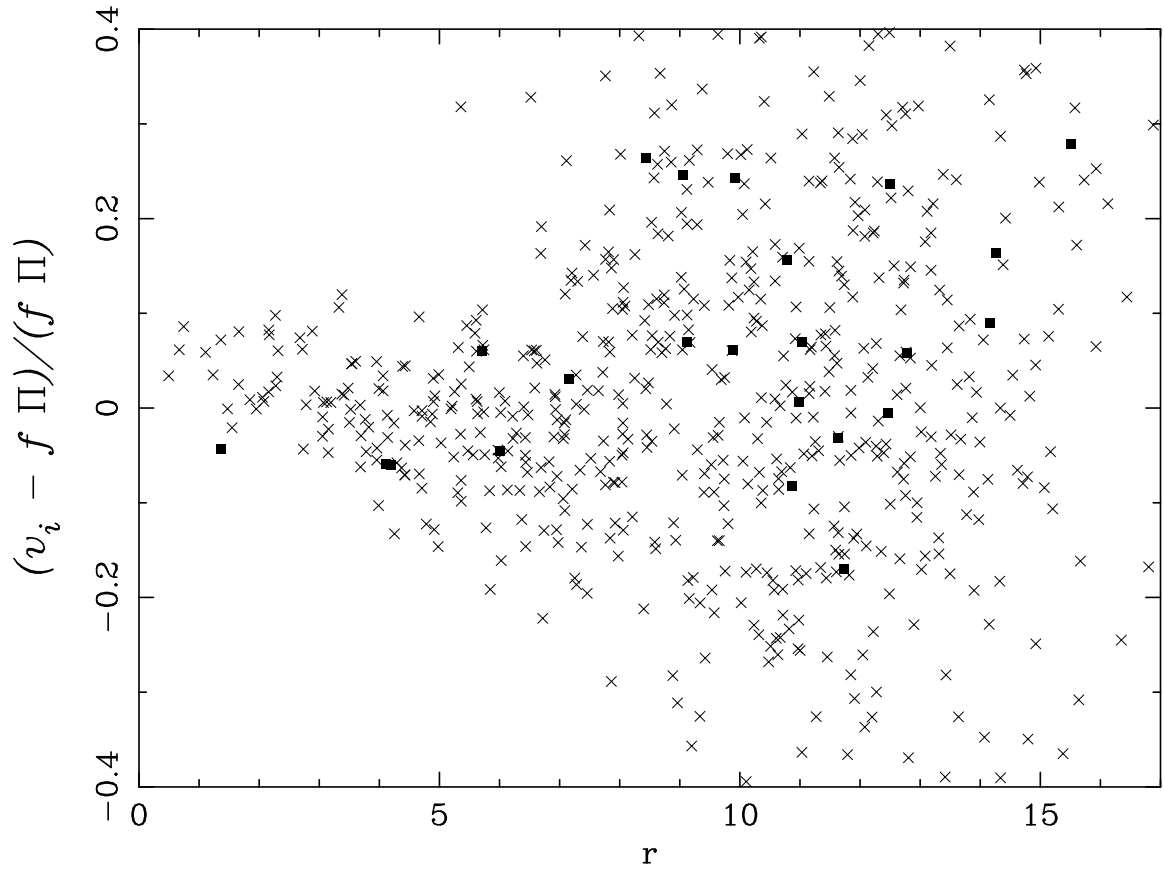}
\caption{c) The same as a) for the FOR~J$0240-3438$ field.}
\end{figure}

\newpage
\begin{figure}[p]
\centering
\includegraphics[angle=-90,scale=0.7]{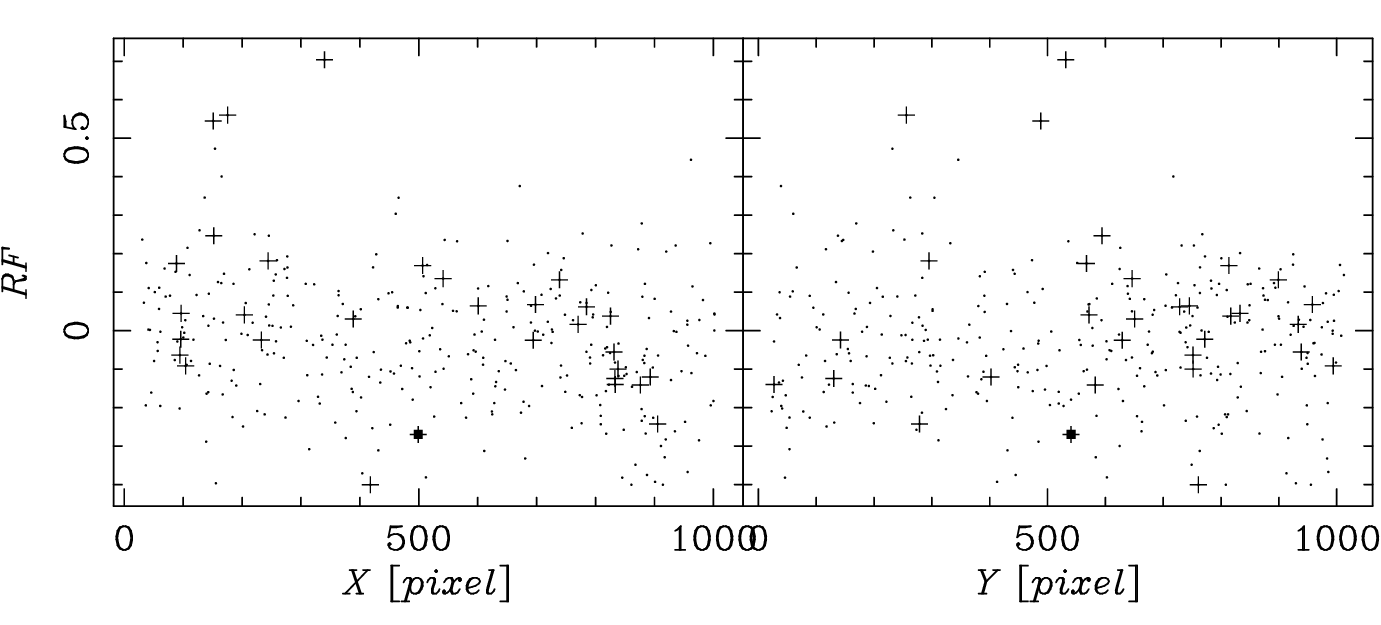}
\includegraphics[angle=-90,scale=0.7]{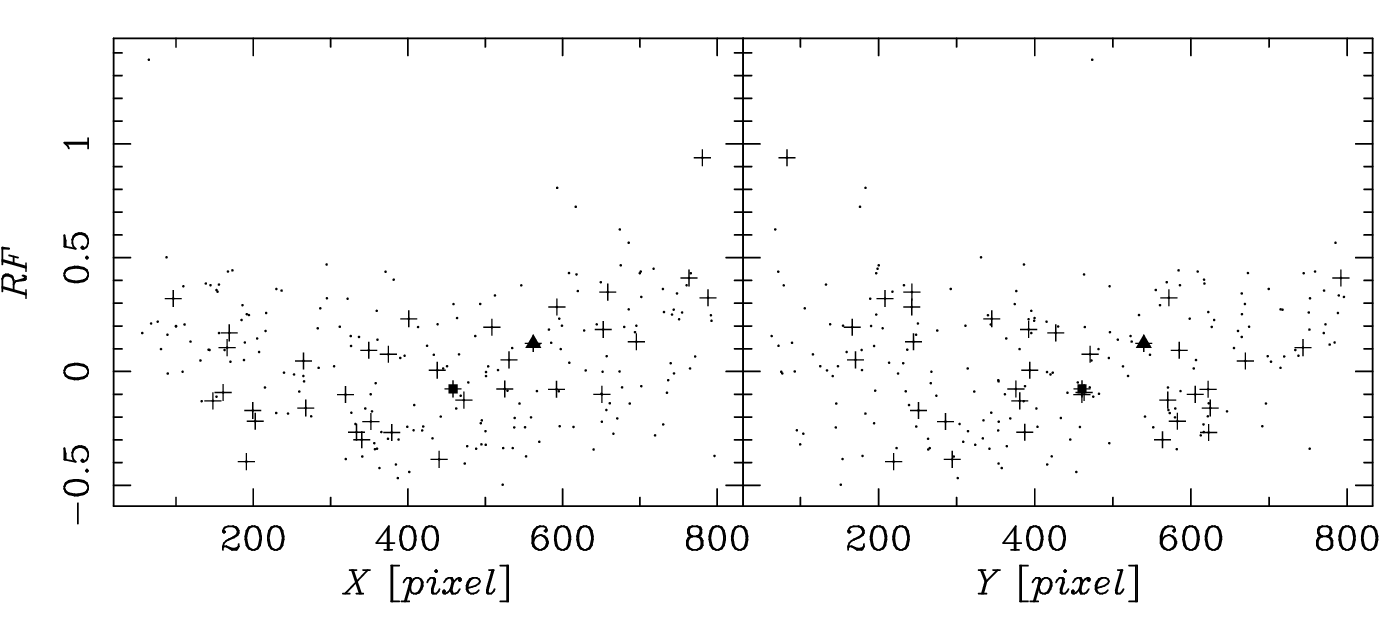}
\caption{a) The left and right panels plot the mean flux residual,
${\cal RF}$, for each object in the fiducial STIS image as a function
of $X_{0}$ and $Y_{0}$, respectively.  The solid square is the QSO and
the pluses are the objects contributing to the construction of the
ePSF (which include the QSO for this image).  b) The same plots for
the fiducial PC image.  The solid square and triangle correspond to
images A and B, respectively, of the lensed QSO.}
\label{fluxres}
\end{figure}

\clearpage
\begin{figure}[p]
\centering
\includegraphics[angle=-90,scale=1.0]{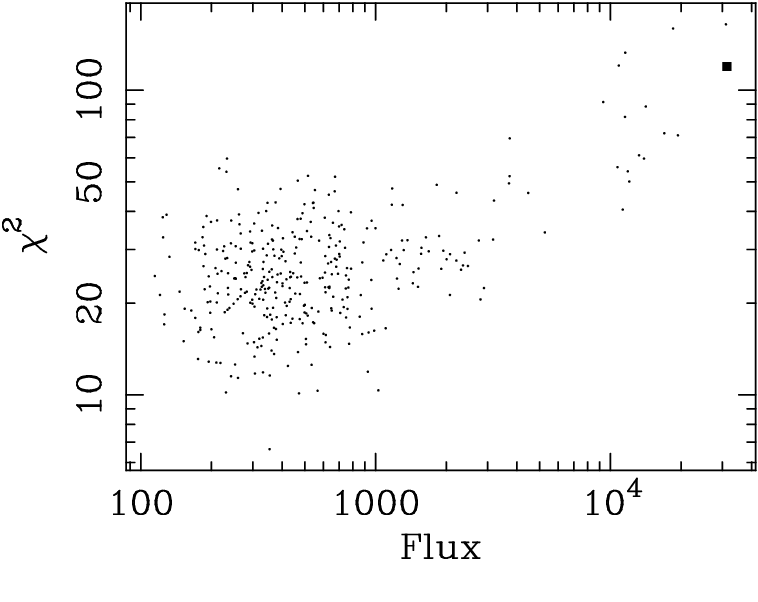}
\includegraphics[angle=-90,scale=1.0]{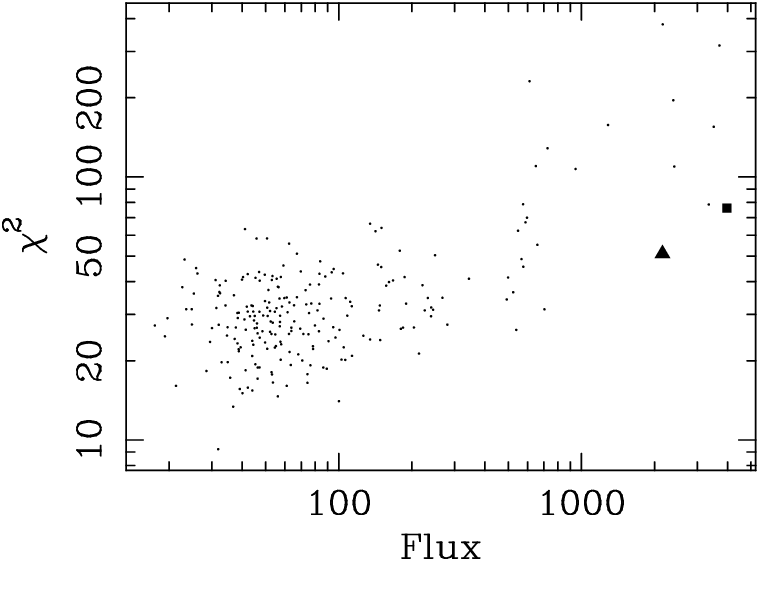}
\caption{a) Plot of $\chi^2$ \textit{versus} total flux for each
object in the fiducial STIS image.  The solid square is the QSO.  b)
The same plot for the fiducial PC image.  The solid square and triangle
correspond to images A and B, respectively, of the lensed QSO.}
\label{fluxchi}
\end{figure}

\clearpage
\begin{figure}[p]
\centering
\includegraphics[angle=-90,scale=0.8]{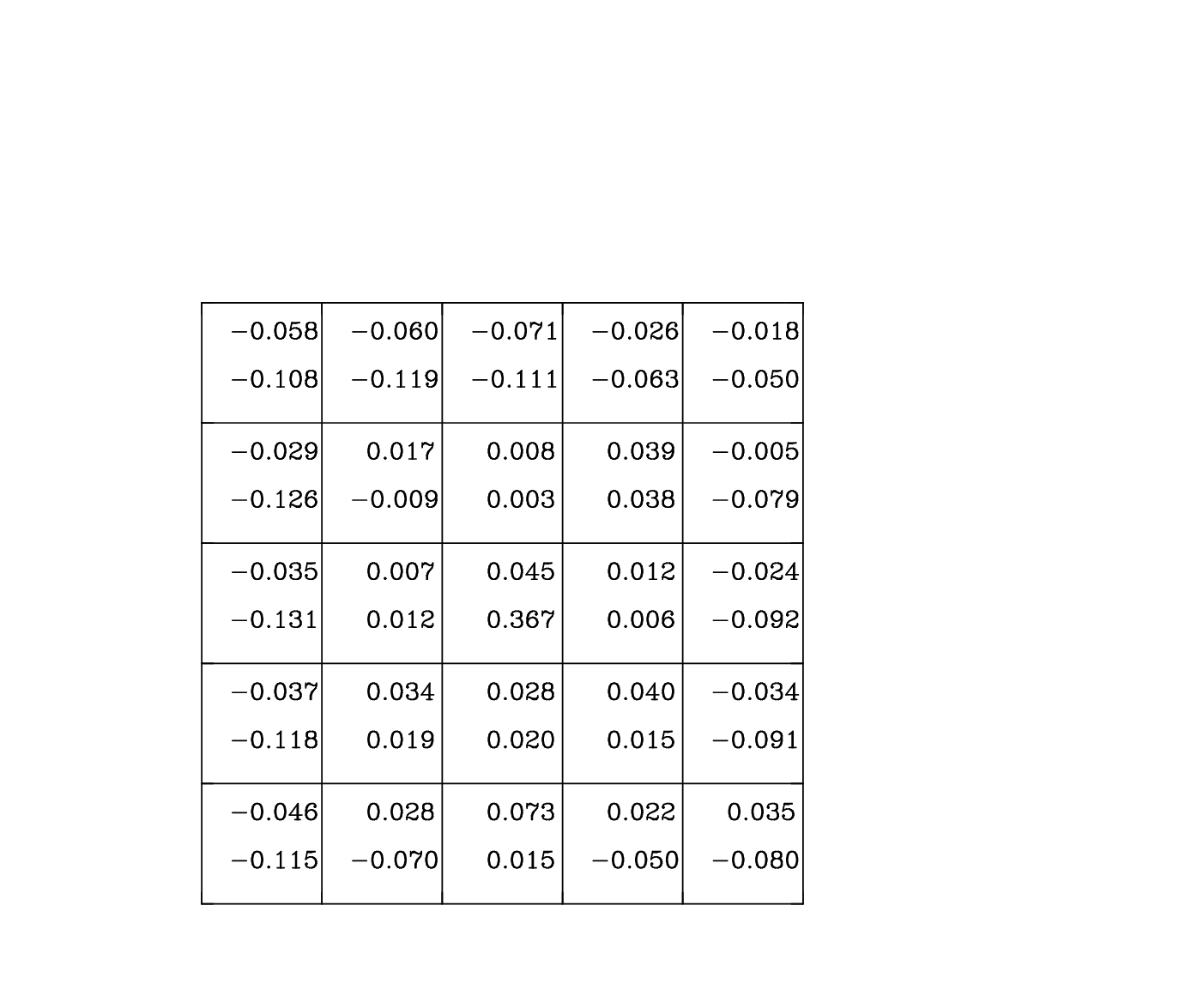}
\caption{The mean flux residual as a function of position within the
science data array for the fiducial STIS image.  The top number in a
square tile is the mean over all objects of the fractional difference
between the value of the science data pixel and the fitted value.  The
bottom number is the mean of the difference between the value of the
science data pixel and the fitted value divided by the uncertainty in
the pixel value.}
\label{numbingrid}
\end{figure}

\clearpage
\begin{figure}[p]
\centering
\includegraphics[angle=-90,scale=0.65]{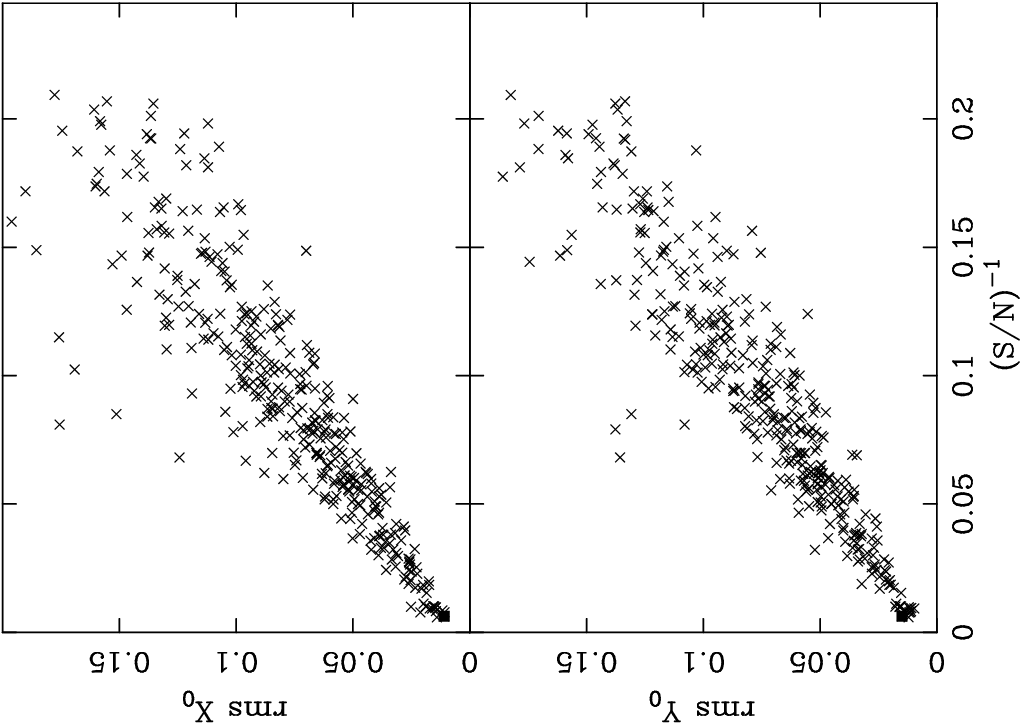}
\includegraphics[angle=-90,scale=0.65]{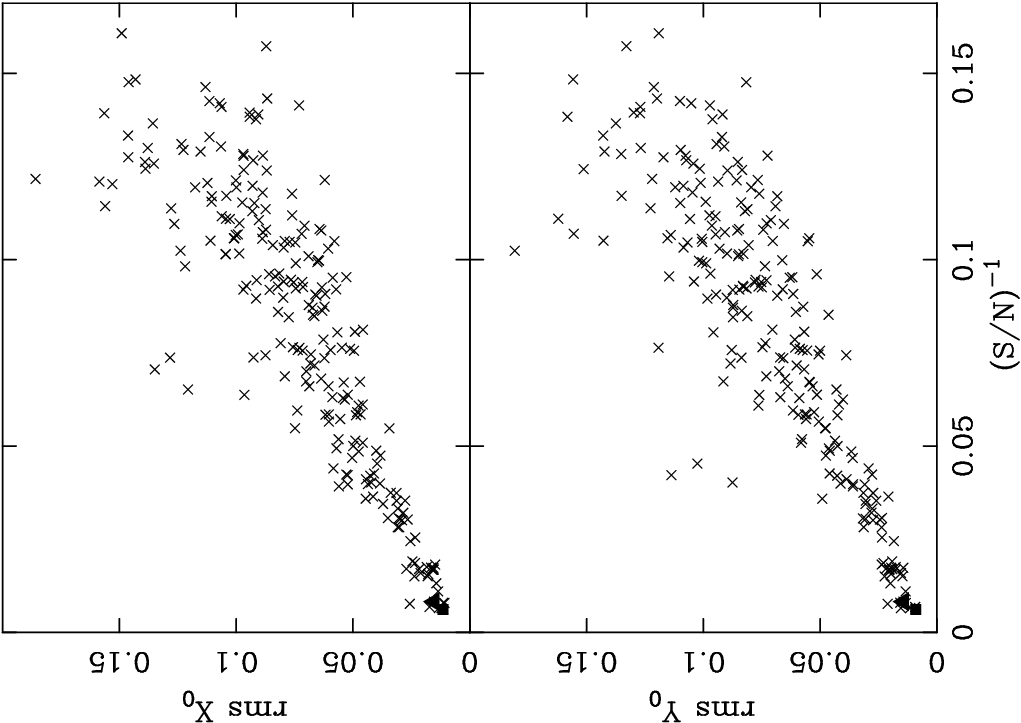}
\caption{a) The \textit{rms} scatter of $X_0$ (top) and $Y_0$
(bottom) around their means \textit{versus} the inverse of the $S/N$
for all objects with at least two measured centroids in the
first-epoch images taken with STIS for the FOR~J$0238-3443$ field.
The solid square is the QSO.  b) The same as (a) for objects in the
first-epoch images taken with the PC for the FOR~J$0240-3434$ field.
The solid triangle is the second image of the lensed QSO.}
\label{rmssn}
\end{figure}

\begin{figure}[p]
\centering
\includegraphics[angle=-90,scale=0.65]{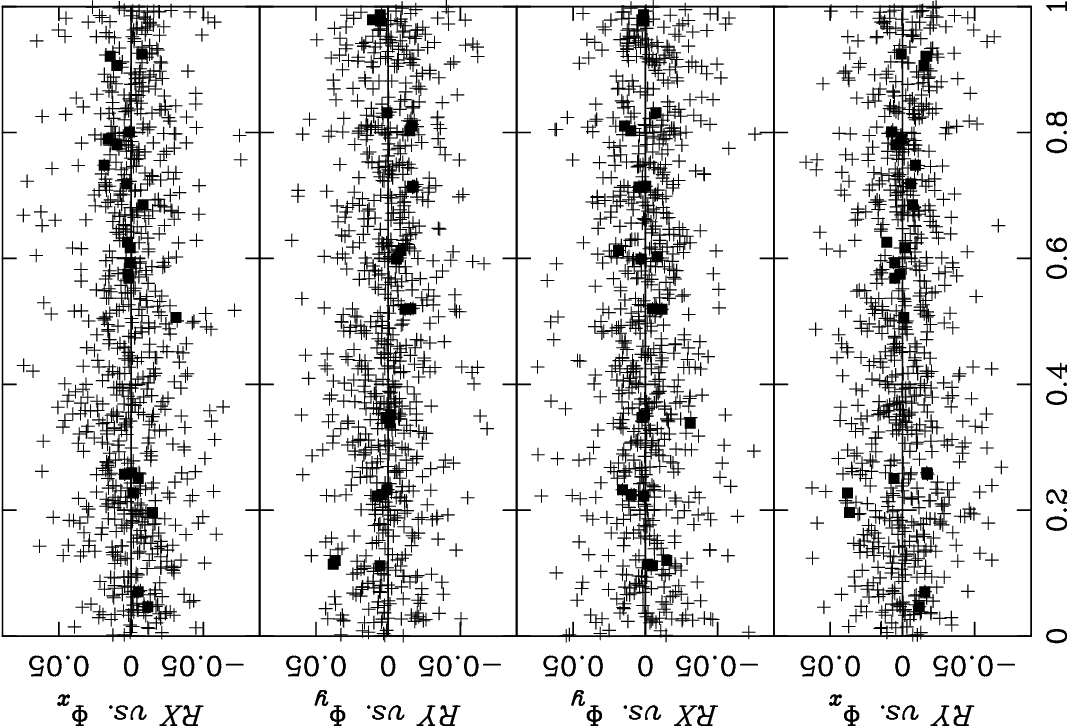}
\includegraphics[angle=-90,scale=0.65]{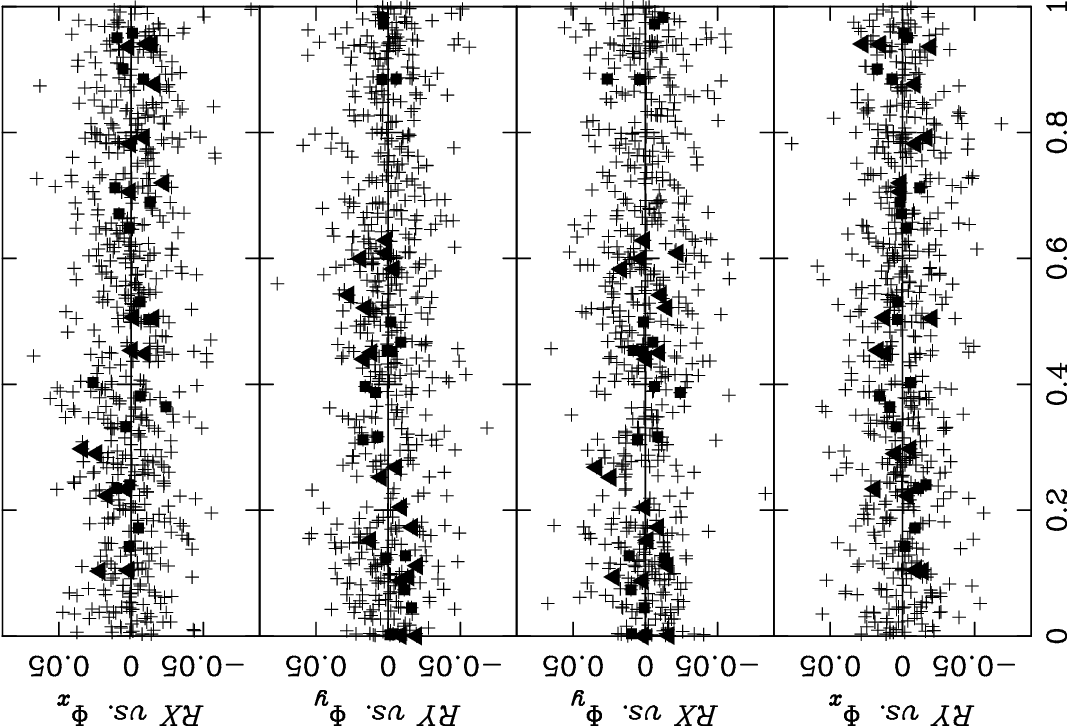}
\caption{a) From top to bottom, the four panels show ${\cal RX}$
\textit{versus} the pixel phase $\Phi_x$, ${\cal RY}$ \textit{versus}
the pixel phase $\Phi_y$, ${\cal RX}$ \textit{versus} $\Phi_y$, and
${\cal RY}$ \textit{versus} $\Phi_x$.  All of the panels show the
objects contributing to the construction of the ePSF for the
first-epoch images taken with STIS for the FOR~J$0238-3443$ field.
The solid square is the QSO.  b) The same as (a) for objects in the
first-epoch images taken with the PC for the FOR~J$0240-3434$ field.
The solid triangle is the second image of the lensed QSO.}
\label{posresid}
\end{figure}

\begin{figure}[p]
\centering
\includegraphics[angle=-90,scale=1.0]{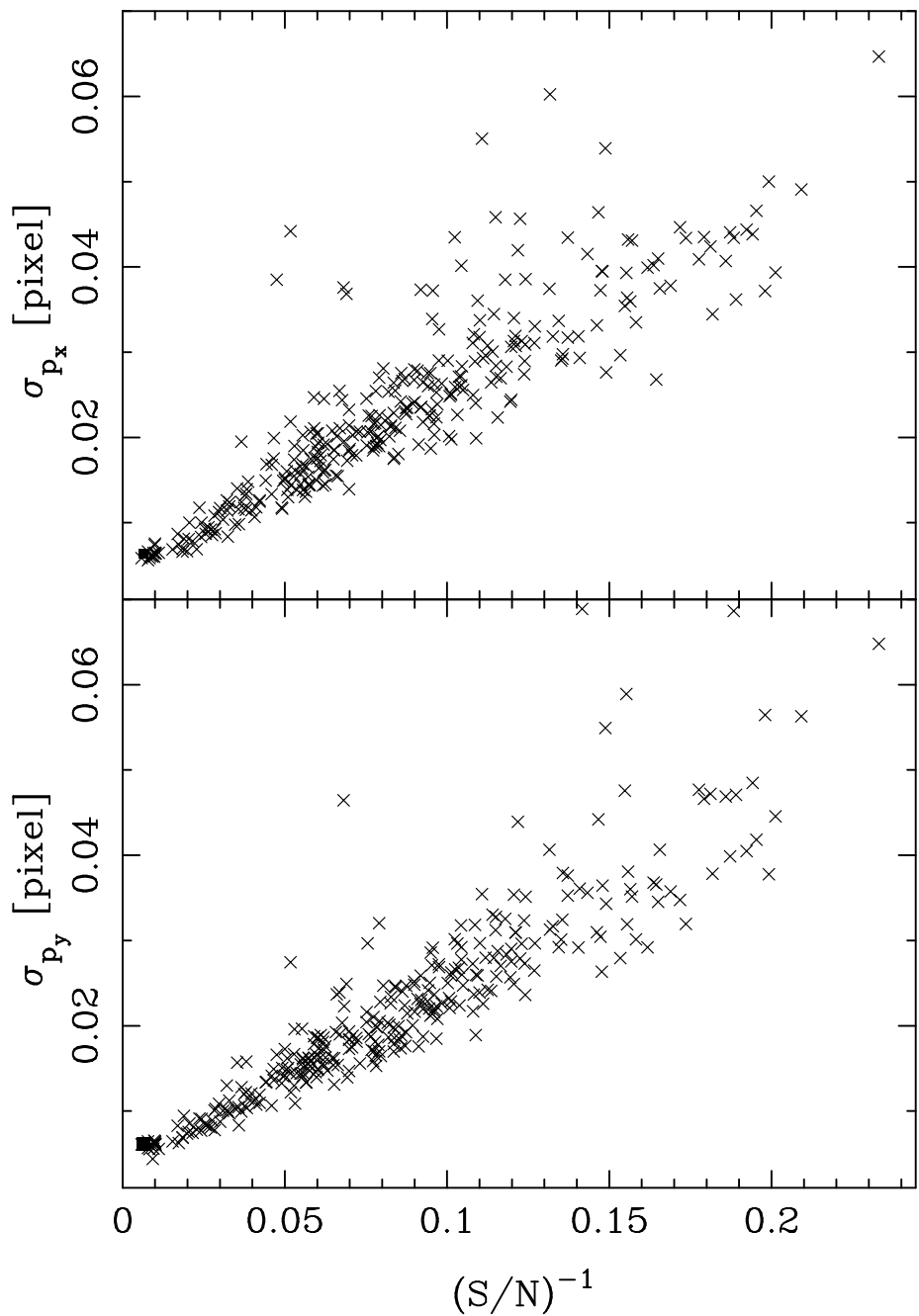}
\caption{The uncertainty in $p_{x}$, $\sigma_{p_x}$, (top) and the
uncertainty in $p_{y}$, $\sigma_{p_y}$, (bottom) plotted
\textit{versus} the inverse of the average $S/N$.  The data are from
epoch 2000 and 2001 for the FOR~J$0238-3443$ field.  The $S/N$ is from
the earlier epoch.  The solid square is the QSO.}
\label{sigvssn}
\end{figure}

\begin{figure}[p]
\centering
\includegraphics[angle=-90,scale=0.7]{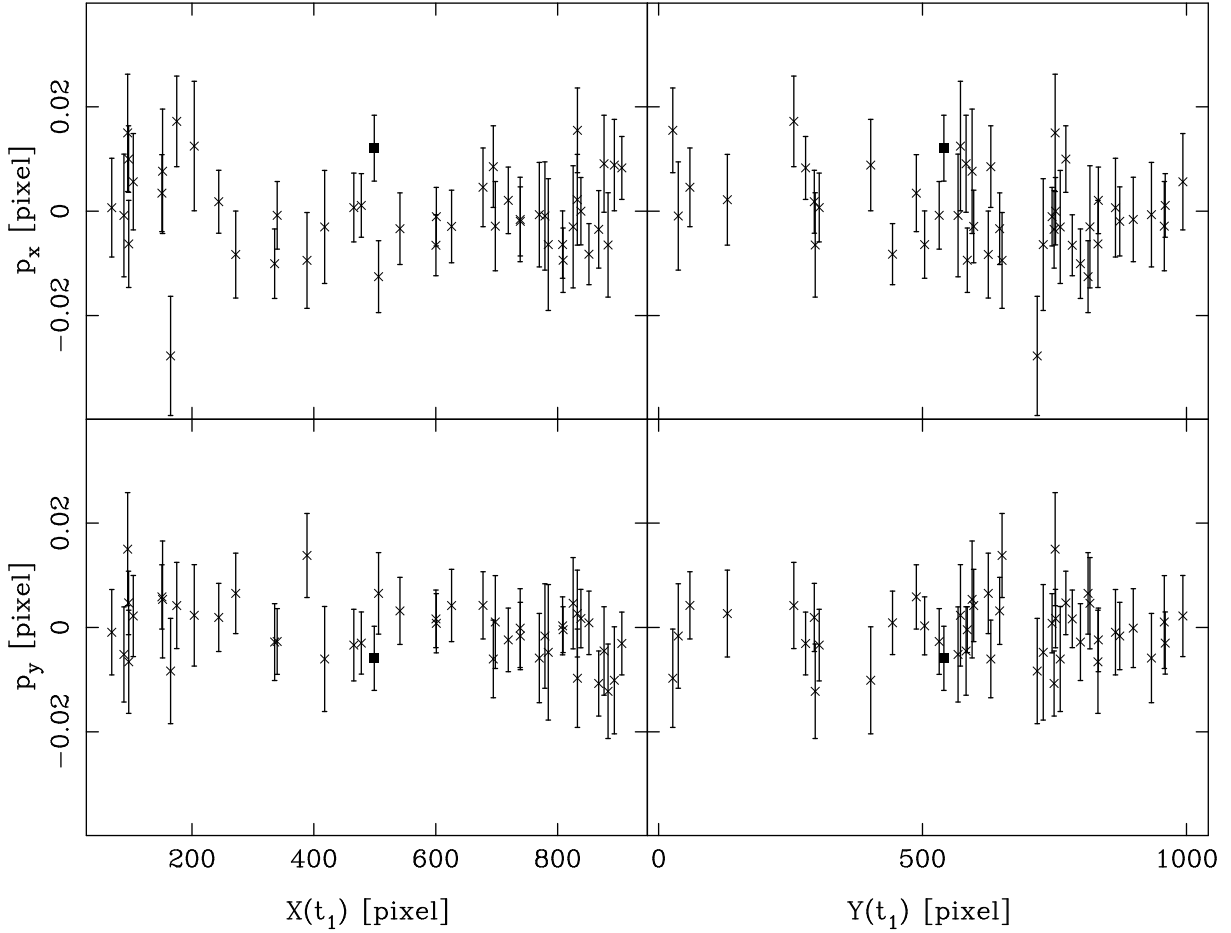}
\caption{The components of the difference in the two centroids of an
individual object in the standard coordinate system, $p_{x}$ and
$p_{y}$, plotted \textit{versus} location in the image for the
FOR~J$0238-3443$ field.  The two centroids for an object are from
epoch 2000 and 2001.  Only objects with $S/N > 30$ are shown.  The
solid square is the QSO.  The left-hand panels plot $p_x$ (top) and
$p_y$ (bottom) \textit{versus} $X(t_{1})$ --- the $X$ coordinate in
the epoch 2000 fiducial image --- and the right-hand panels similarly
plot the same quantities \textit{versus} $Y(t_{1})$ --- the $Y$
coordinate in the epoch 2000 fiducial image.}
\label{pxpyxy}
\end{figure}

\begin{figure}[p]
\centering
\includegraphics[angle=-90,scale=0.9]{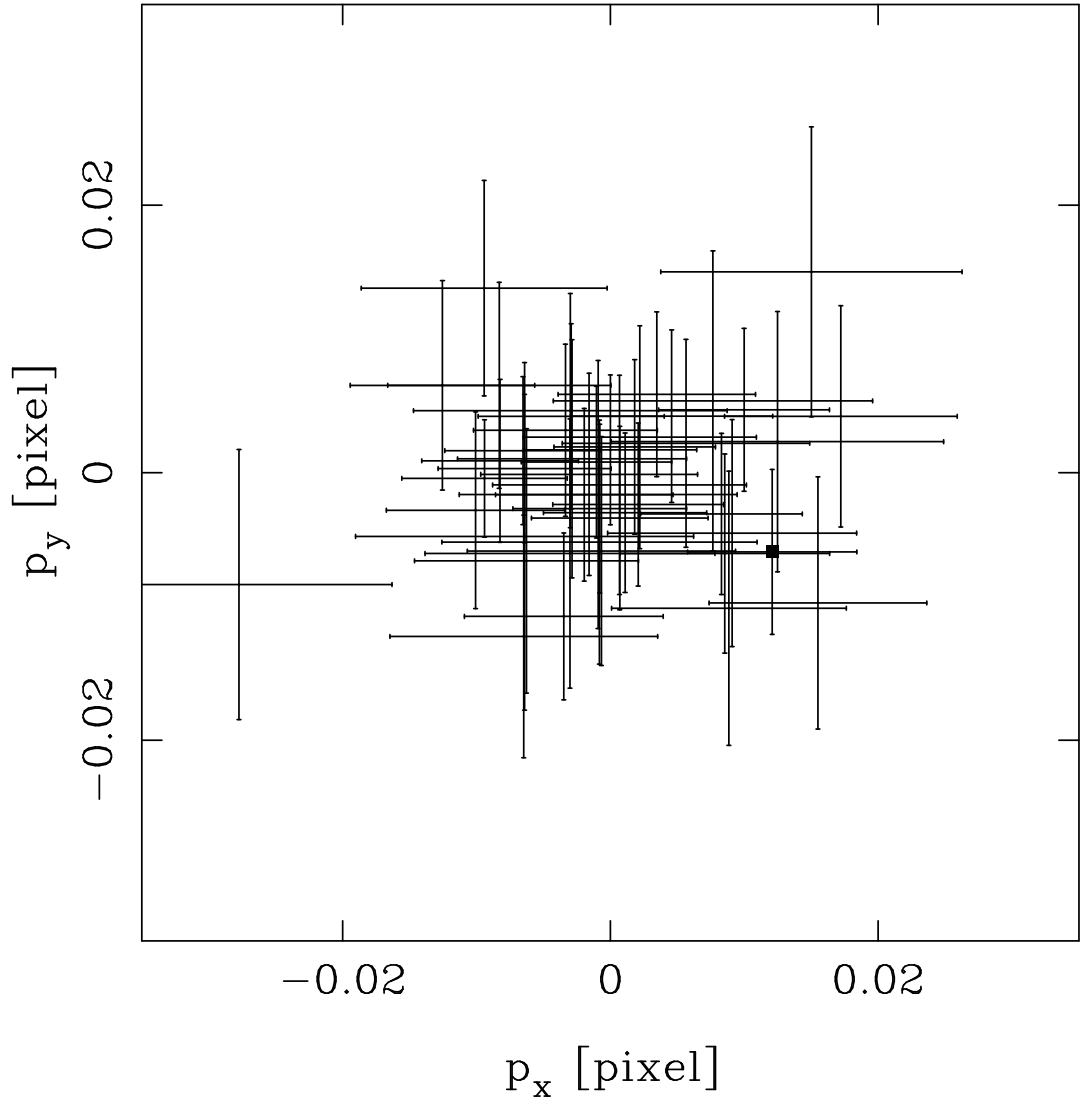}
\caption{Plot of $p_{y}$ {\it versus} $p_{x}$ for the 45 objects with
$S/N > 30$ in the FOR~J$0238-3443$ field using epoch 2000 and 2001
data.  The solid square is the QSO.}
\label{pxvspy}
\end{figure}

\begin{figure}[p]
\centering
\includegraphics[angle=-90,scale=0.9]{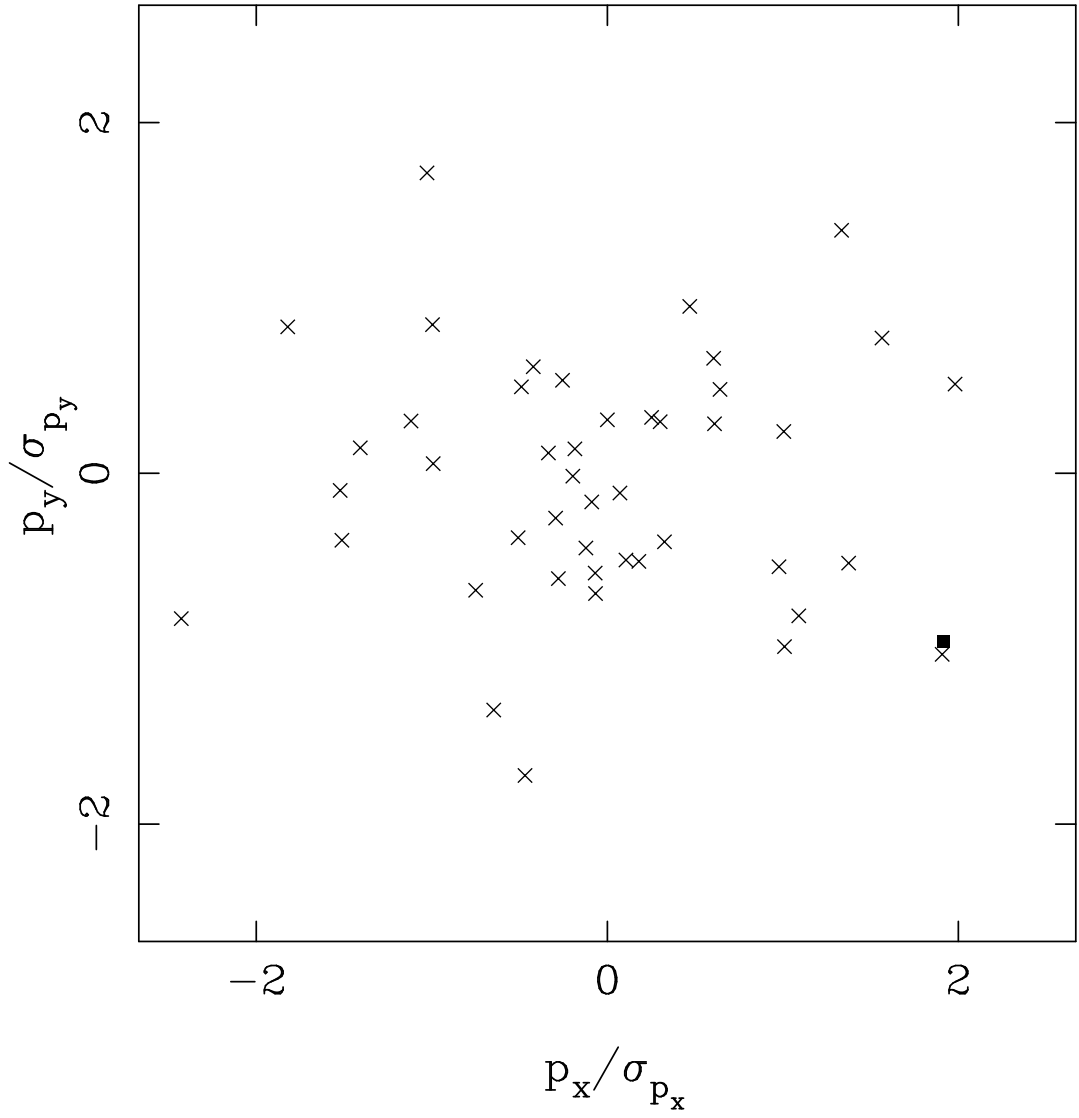}
\caption{Plot of $p_{y}/\sigma_{p_{y}}$ {\it versus}
$p_{x}/\sigma_{p_{x}}$ for the same objects as in Figure~\ref{pxvspy}.}
\label{pxvspy2}
\end{figure}


\clearpage
\begin{references}
\reference{} Anderson, J. \& King, I. R. 1999, PASP, 111, 1095
\reference{} Anderson, J. \& King, I. R. 2000, PASP, 112, 1360
\reference{} Baggett, S. et al. 2002, in HST WFPC2 Data Handbook, v 4.0,
ed. B. Mobasher, Baltimore, STScI
\reference{} Bedin, L. R., Anderson, J., King, I. R., \& Piotto, G. 2001,
ApJ, 560, L75
\reference{} Bernstein, G. 2002, PASP, 114, 98
\reference{} Brown, T., et al. 2002, in HST STIS Data Handbook, version
4.0, ed. B. Mobasher, Baltimore, STScI
\reference{} Cudworth, K. M., Olszewski, E. W., \& Schommer, R. A. 1986,
AJ, 92, 766
\reference{} Dehnen, W., \& Binney, J. J. 1998, MNRAS, 298, 387
\reference{} King, I. R., Anderson, J., Cool, A. M., \& Piotto, G. 1998,
ApJ, 492, L37
\reference{} Kuijken, K., \& Rich, R. M. 2002, astro-ph/0207116
\reference{} Kunkel, D. E., \& Demers, S. 1977, ApJ, 214, 21
\reference{} Lauer, T. R. 1999, PASP, 111, 1434
\reference{} Lynden-Bell, D. 1976, MNRAS, 174, 695
\reference{} Lynden-Bell, D. 1982, Observatory, 102, 202
\reference{} Lynden-Bell, D. 1983, in Internal Kinematics and Dynamics of
Galaxies, IAU Symp.~100, ed. E. Athanassoula (Dordrecht: Reidel), 89
\reference{} Lynden-Bell, D. 1994, in Dwarf Galaxies, ESO/OHP Workshop
Proceedings No.~49, ed. G. Meylan \& P. Prugniel (Garching: ESO), 589
\reference{} Lynden-Bell, D., \& Lynden-Bell, R. M. 1995, MNRAS, 275, 429
\reference{} Majewski, S. R. 1994, ApJ, 431, L17
\reference{} Mateo, M. 1998, ARAA, 36, 435
\reference{} Monet, D. G.; Dahn, C. C., Vrba, F. J., Harris, H. C.,
Pier, J. R., Luginbuhl, C. B., \& Ables, H. D. 1992, AJ, 103, 638
\reference{} Scholtz, R. -D. \& Irwin, M. J. 1993 in IAU Symp. 161,
Astronomy from Wide-Field Imaging, ed. H. T. MacGillivray et
al. (Dordrecht: Kluwer), p. 535
\reference{} Press, W. H., Teukolsky, S. A., Vetterling, W. T., \&
Flannery, B. P. 1992, Numerical Recipes (Cambridge, Cambridge Univ.
Press)
\reference{}Shaklan, S., Sharman, M. C., \& Pravdo, S. H. 1995,
Appl. Opt., 34, 6672
\reference{} Stetson, P. B. 1987, PASP, 99, 191 
\reference{} Stetson, P. B. 1992, in ASP Conf. Ser. Vol. 25,
Astronomical Data Analysis Software and Systems, ed. D. M. Worrall, C. 
Biemesderfer, \& J. Barnes (San Francisco: ASP), 297  
\reference{} Stetson, P. B. 1994, PASP, 106, 250
\reference{} Schweitzer, A. E., Cudworth, K. M., \& Majewski,
S. R. 1997, in ASP Conf. Ser. 127, Proper Motions and Galactic
Astronomy, ed. R. M. Humphreys (San Francisco: ASP), 132
\reference{} Smart, W. M. 1997, Textbook on Spherical Astronomy
(Cambridge: Cambridge University Press)
\reference{} Tinney, C. G. 1995, MNRAS, 277, 609
\reference{} Tinney, C. G. 1996, MNRAS, 281, 644
\reference{} Tinney, C. G., Da Costa, G. S., \& Zinnecker, H. 1997,
MNRAS, 285, 111
\reference{} Tinney, G. G., Reid, I. N., Gizis, J., \& Mould, J. R. 1995,
AJ, 110, 3014
\reference{} van den Bergh, S. 2000, The Galaxies of the Local Group,
Cambridge Astrophysics Series No. 35 (Cambridge: Cambridge University Press)
\end{references}
\end{document}